\documentclass[aps.prb,11pt]{revtex4-2}

\usepackage{amsmath,amssymb,mathrsfs}

\usepackage{graphicx}   
\usepackage{verbatim}   
\usepackage{color}      
\newcommand{\norm}[1]{\left\lVert#1\right\rVert}
\newcommand{\editor}[2]{%
  \expandafter\newcommand\csname #1note\endcsname[1]{%
    \textcolor{#2}{(\textbf{#1:} ##1)}}%
  \expandafter\newcommand\csname #1\endcsname[1]{%
    \textcolor{#2}{##1}}%
  \expandafter\newcommand\csname #1cancel\endcsname[1]{%
    \textcolor{#2}{\sout{##1}}}%
  \expandafter\newcommand\csname #1change\endcsname[2]{%
    \textcolor{#2}{\sout{##1} ##2}}%
  \newenvironment{#1text}{\color{#2}}{\color{black}}
}
\editor{V}{magenta} 

\begin{document}


\title{Understanding first order Raman spectra of boron carbides across the homogeneity range}

\author{Guido Roma}
\email{guido.roma@cea.fr}
\author{Kevin Gillet}
\affiliation{Universit\'e Paris-Saclay, CEA, Service de Recherches de M\'etallurgie Physique, 91191 Gif sur Yvette, France}

\author{Antoine Jay}
\affiliation{Laboratoire d'analyse et d'architecture des syst\`emes, CNRS, 31031 Toulouse c\'edex 4, France}

\author{Nathalie Vast}
\affiliation{Laboratoire des Solides Irradi\'es, CEA/DRF/IRAMIS, CNRS, \'Ecole Polytechnique, Institut Polytechnique de Paris,
91120 Palaiseau, France}

\author{Ga\"elle Gutierrez}
\affiliation{Universit\'e Paris-Saclay, CEA, Laboratoire Jannus, Service de Recherches de M\'etallurgie Physique, 91191 Gif sur Yvette, France}

\begin{abstract}
Boron carbide, a lightweight, high temperature material, has various applications as a structural material and as a neutron absorber. 
The large solubility range of carbon in boron, between $\approx$ 9~\% and 20~\%, has been theoretically explained by some of us by the thermodynamical stability of
three icosahedral phases at low temperature, with respective carbon atomic concentrations: 8.7~\% (B$_{10.5}$C, named OPO$_1$), 13.0~\% (B$_{6.7}$C, named OPO$_2$), whose theoretical Raman spectra are still unknown, 
and 20~\% (B$_4$C), from which the nature of some of the Raman peaks are still debated.
We report theoretical and experimental results of the first order, non-resonant, Raman spectrum of boron carbide. 
Density functional perturbation theory enables us to obtain the Raman spectra of the OPO$_1$ and OPO$_2$ phases, 
which are perfectly ordered structures with however a complex crystalline motif of 414~atoms, due to charge compensation effects. 
Moreover, for the carbon-rich B$_4$C, with a simpler 15-atom unit cell, we study the influence of the low energy point defects and of their concentrations on the Raman spectrum, in connection with experiments, thus providing insights into the sensitivity of experimental spectra to sample preparation, experimental conditions and setup. 
In particular, this enables us to propose a new structure at 19.2~\% atomic carbon concentration, B$_{4.2}$C, that, within the local density approximation of density functional theory (DFT-LDA), lies very close to the convex hull of boron carbide, on the carbon-rich side. 
This new phase, derived from what we name the ``3+1'' defect complex, helps in reconciling the experimentally observed Raman spectrum with the theory around 1000~cm$^{-1}$.
Finally, we  predict the intensity variations induced by the experimental geometry
and quantitavely assess the localisation of bulk and defect vibrational modes and their character, with an analysis of ``chain'' and ``icosahedral'' modes. 

\end{abstract}

\maketitle

\section{Introduction}
Boron carbide is a ceramic material that combines a number of practical applications, interesting structural peculiarities, shared with other boron based compounds, and prospective applications which are still far from their realization.
Due to its hardness it is largely used for ballistic protection and for abrasive blast nozzle, due to its wear resistance. The non negligible percentage of the  $^{10}$B isotope makes it useful as neutron absorber, a sort of ``nuclear functional material''~\cite{banerjee_10_2012}, used to control and/or stop the chain reaction in nuclear power plants.
As far as  
its electronic properties are concerned, boron carbide is a semiconductor; as such, various functional applications have been envisaged, for example in the field of thermoelectricity, already decades ago~\cite{Bouchacourt_correlation_1985} or as catalyst for photoelectrochemical water splitting~\cite{liu_boron_2013}, but the difficulties in controlling the stoichiometry and the microstructure have up to now hindered such uses.


The electronic band gap, a crucial quantity for functional applications, is still somewhat controversial, although experiments suggest it is approximately between 2.1 and 2.4~eV~\cite{werheit_on_2006,hushur_high-pressure_2016}. Theoretical estimates based on usual semilocal density functional theory (DFT) (a method which is known to typically underestimate band gaps for most materials) for ideal carbon-rich B$_4$C stoichiometry give larger band gap values (around 2.8--3~eV). For the same structure, denoted (B$_{11}$C)C-B-C ---where (B$_{11}$C) stands for 12-atom icosahedra linked through 3-atom C-B-C chains--- more accurate theoretical estimations obtained with hybrid functionals provide, as expected, even larger values, on the order of 3.6-3.8 eV~\cite{rasim_local_2018,gillet_influence_2018}. It is then clear that the experimental results are probing electronic levels of defective structures~\cite{werheit_optical_1992,werheit_assignment_2018}.  As far as the boron-rich phases are concerned, the ideal B$_{13}$C$_2$ structure based on (B$_{12}$) icosahedra and C-B-C chains, is metallic in DFT, in clear disagreement with experiment. 
Some local arrangements have been proposed with defective chains and boron interstitials~\cite{rasim_local_2018,ektarawong_structural_2018}, whose electronic structure is compatible with the semiconducting character of the material.

However, two new phases, B$_{10.5}$C and B$_{6.7}$C, denoted OPO$_1$ and OPO$_2$ at respectively 8.7~\% and 13.0~\%  atomic carbon concentrations,  also preserve the semiconducting properties~\cite{jay_theoretical_2019}. They have been shown to be the thermodynamically stable phases in the boron-rich domain, bringing  the theoretical solubility range of carbon in boron close to the experimental one in DFT with the generalized gradient approximation (GGA).  In the present work, we report the Raman spectra of the ordered OPO$_1$ and OPO$_2$ phases, and introduce a polymorph of B$_{6.7}$C, and name this variant of OPO$_2$ as OPO$_2^{panto}$.

Experimentally, boron carbide is observed to have the rhombohedral $R\overline{3}m$ symmetry. 
On the theoretical side, the OPO$_1$ and OPO$_2$ phases have the lowest formation energy at fixed stoichiometry and define the convex hull of boron carbides in the
boron-rich domain. They have the  $P$1 triclinic symmetry and can be represented on average by a $R\overline{3}m$ structure~\cite{jay_theoretical_2019}. The atomic structure consists of  (B$_{12}$) icosahedra connected by three-atom C-B-C chains, however, because of charge compensation, replacing one third of the C-B-C chains with either~B$_4$
blocks (OPO$_1$) ---where 2 of the boron atoms are in a $6g$ extra interstitial Wyckoff site (EIWS) of the $R\overline{3}m$ unit-cell~\cite{jay_theoretical_2019}--- or with chain-like~C-B-C$\cdots$B blocks ( OPO$_2$) ---where the extra B-atom is in a $2c$ EIWS~\cite{jay_theoretical_2019}--- is energetically favorable.  The B$_4$ intericosahedral arrangements have been reported in previous works for other carbon concentrations, both theoretically~\cite{Shirai:2014,rasim_local_2018} and experimentally~\cite{Yakel:1975}. The latter experimental report has however not been subsequently confirmed.

Both the OPO$_1$ and OPO$_2$ structures are perfectly ordered with a crystalline motif of 414~atoms, and some of us have proposed that the 15-atom unit cell with the $R\overline{3}m$ symmetry observed in x-ray and neutron experiments is the averaged value of the OPO$_1$ (or OPO$_2$) phase~\cite{jay_theoretical_2019}. Such an average by experiment of an otherwise complex theoretical motif (which has an energy lower than the averaged structure),  has been found for other materials, for instance in Refs.~\onlinecite{wang-2020,zunger-2021}.  In boron carbides, the averaging implies an  ordered partial occupation (OPO) of some specific interstitial Wyckoff crystallographic sites of the 15-atom unit cell, which was part of the reason for which boron carbide was considered as an "intrinsically" disordered material~\cite{rasim_local_2018,widom-2012,werheit_assignment_2018}. 

On the carbon-rich side, B$_4$C corresponds to an atomic arrangement based on icosahedra connected by three-atoms C-B-C chains. 
The carbon-rich structure shows (B$_{11}$C) icosahedra. 
This ideal structure is probably not exactly corresponding to the real crystalline arrangements, but is the scaffold on which the inclusion of various defects leads to the actual structure of boron carbide around 20~\% atomic carbon concentration. 

The attainability of the B$_4$C stoichiometry without carbon segregation is still a subject of debate, the highest carbon concentration experimentally observed being  B$_{4.3}$C, whereas the ideal (theoretical) B$_4$C structure, (B$_{11}$C)C-B-C, is  quasi-rhombohedral, with a slight monoclinic distortion due to icosahedral carbon, which breaks the inversion symmetry of the $R\overline{3}m$ space group, leading to the $C_m$ symmetry.

In the lowest energy configuration of this kind, the icosahedral carbon is one of the polar atoms~\cite{lazzari_atomic_1999}, or (B$_{11}$C$^p$)C-B-C; this is commonly accepted as the structure which most closely corresponds to experimental realisations of carbon-rich boron carbide, as also confirmed by substantial agreement of the calculated vibrational modes with the experimental Raman spectrum~\cite{lazzari_atomic_1999,vast_vibrational_2000,domnich_boron_2011}. From now on, when discussing the (B$_{11}$C)C-B-C structure, we intend (B$_{11}$C$^p$)C-B-C.

Several defective structures have been put forward as possibly present in carbon-rich boron carbide. C-B-B chain variants have been suggested for boron-rich~\cite{werheit_carbon-insertion_1994} and also for nominal B$_4$C compositions~\cite{kwei_structures_1996}. Other chain variants were also proposed. In particular, C-$\square$-C arrangements  (corresponding to a boron chain vacancy) and C-C chains (corresponding to a C-C bond formation) were studied with some detail in the neutral state~\cite{betranhandy-2012,raucoules_mechanical_2011,schneider_stability_2017,you_first-principles_2018}, and more recently also in charged states~\cite{gillet_influence_2018}. Other authors have investigated the interplay of various forms of disorder, including various antisites and vacancies~\cite{huhn_free-energy_2013,ektarawong_first-principles_2014,ektarawong_configurational_2015,ektarawong_structural_2018}.

\begin{figure}
\includegraphics[width=0.7\columnwidth]{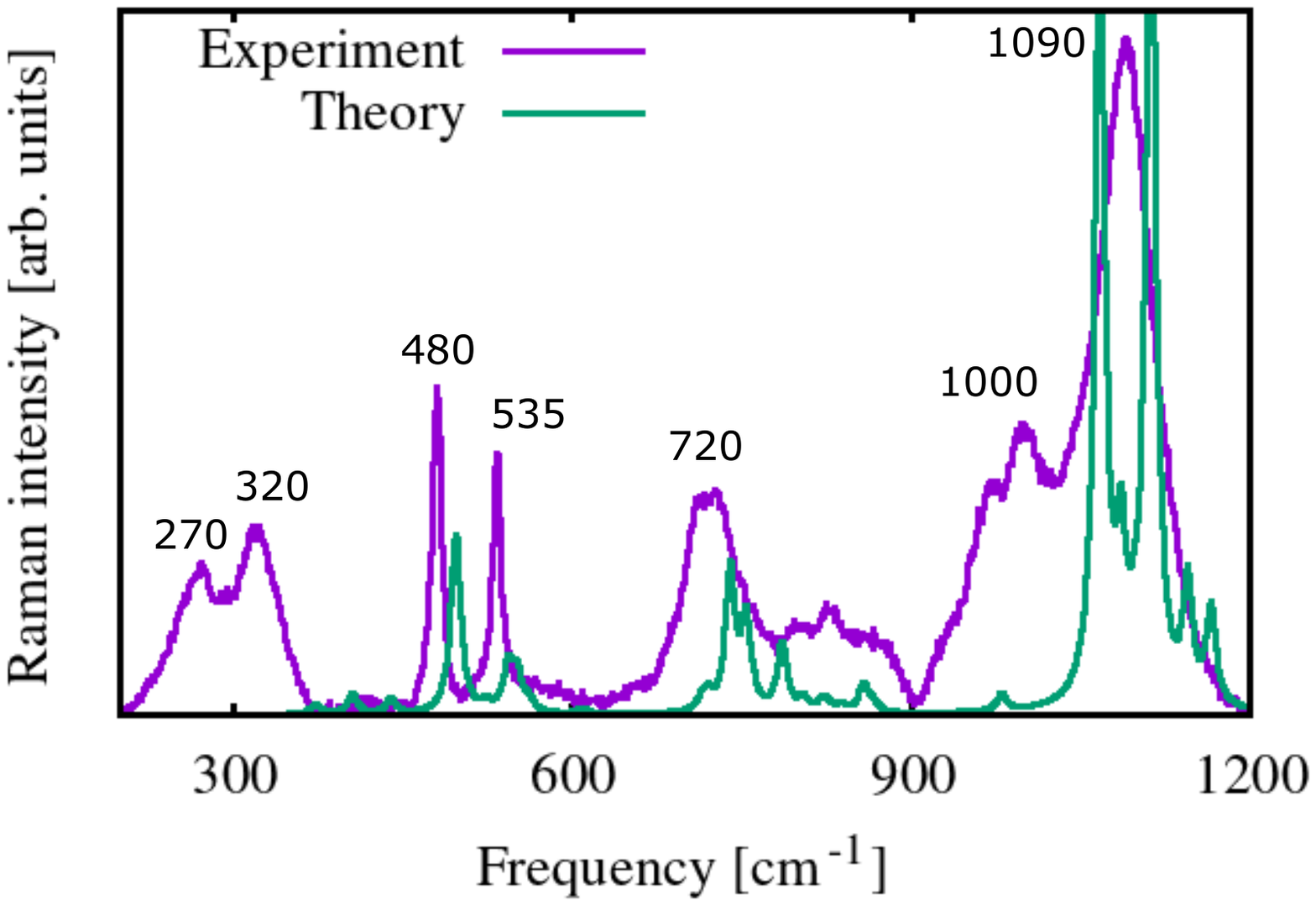}
\caption{Typical experimental Raman spectrum of carbon-rich boron carbide compared to the theoretical first order Raman spectrum of the (B$_{11}$C)C-B-C structure. The experimental one presented here has been obtained on a sample with small grains and is thus randomized
over light polarization directions, as the theoretical one, with an incident laser with wavelength of 633~nm. The frequencies of the main experimental peaks are reported.}
\label{FigExpGrainsFins}
\end{figure}

The influence of chain variants, carbon/boron substitution and, more generally, disorder on the Raman spectrum of boron carbide has already been studied by means of first principles approaches~\cite{jay_carbon-rich_2014,jay_conception_2016,kunka_crystallographic_2016,kunka_evaluating_2017}. However, such structural variations were studied as global modifications of the 15-atom unit-cell of (B$_{11}$C)C-B-C, not as point defects at relatively low concentrations. For example, the Raman spectrum of the (B$_{11}$C)C-C phase was studied in detail~\cite{jay_carbon-rich_2014}, and various possible chain and icosahedral variants were subsequently investigated~\cite{jay_conception_2016,kunka_crystallographic_2016,kunka_evaluating_2017}. The phases which contain chain variants are found to yield entirely different spectra, because they correspond to globally different phases, in general with much higher formation energy with respect to that of (B$_{11}$C)C-B-C; such phases are drastically different from the (B$_{11}$C)C-B-C phase with point defects at relatively low concentration.  It is reasonable to use the spectra of such phases as building blocks of a reconstituted spectrum of a boron carbide sample only if they are present as sufficiently large crystallites; it is, however, less justified if the chain variants (or icosahedral ones) occur as diluted point defects.
In spite of these efforts~\cite{jay_conception_2016,kunka_crystallographic_2016,kunka_evaluating_2017}  some of the features observed in the experimental spectrum, like the one that we show in Figure~\ref{FigExpGrainsFins}, are not fully understood. The main peak in Raman spectra of boron carbide, located at 1080-1100 cm$^{-1}$, is not resolved experimentally, while theoretical calculations show that it originates from two distinct peaks. The shoulder just below, at approximately 1000~cm$^{-1}$ in the experimental spectrum, is explained by none of the considered variants theoretically proposed~\cite{kunka_crystallographic_2016}. The presence of the low frequency doublet around 300~cm$^{-1}$ (Fig.~\ref{FigExpGrainsFins})
strongly depends on the direction of  polarization of the incident light and on pressure~\cite{Yan_B4Cdeamorph_PRL2009,guo_pressure-induced_2010}; an unambiguous theoretical interpretation of this doublet is still missing.

In the present paper we adopt an approach which is slightly different with respect to previous theoretical works devoted to the Raman spectrum of boron carbide: we consider point defects at relatively low concentrations and we investigate their influence on the Raman spectrum of boron carbide.
In section~\ref{Method} we describe the theoretical tools and experimental settings used to obtain the results.

Before discussing the results on point defects, in section~\ref{Geometry} we discuss the influence of the experimental geometry and introduce, in section~\ref{Localisation}, a general measure of mode localisation. The discussion on geometrical effects is complemented by some experimental results on large grains of boron carbide with varying orientations. 

Section~\ref{Defects} is devoted to our findings concerning the influence of point defects on the Raman spectrum of carbon-rich boron carbide. They were obtained thanks to direct calculations of the first order, non-resonant, Raman spectrum for reasonably sized supercells and through an embedding procedure recently described~\cite{roma_modeling_2019}. We conclude this section by presenting the Raman spectrum of defect complexes and of the ordered boron-rich phases recently discovered~\cite{jay_theoretical_2019} and discuss their relationship with the previously presented point defects.




\section{Methods and technical details}
\label{Method}

\subsection{Theoretical methods}
The calculation of first order Raman spectra are performed in the framework of density functional perturbation theory (DFPT) and they rely on the Placzek approximation~\cite{lazzeri_first-principles_2003}. We used the local density approximation (LDA), with a plane wave basis set and norm conserving pseudopotentials. A cutoff of 80 Rydberg was used for all calculations presented here. Most direct defect calculations were performed in a 2$\times$2$\times$2 supercell of the trigonal 15-atom unit cell (120 atoms) of the (B$_{11}$C)C-B-C structure, with a 3$\times$3$\times$3 $\Gamma$-centered {\bf k}-point mesh for the sampling of the supercell Brillouin Zone (BZ) (only 2$\times$2$\times$2 for formation energies reported in Section~\ref{DefectsFormationEnergies} {\bf k}-point mesh). Total energy calculations were performed at the equilibrium volume of the undefected B$_4$C crystal. We stress that this may, in some cases, lead to spurious negative frequencies in the vibrational spectrum. However, as we are not specifically interested in low frequency phonons here, we prefer to avoid the risk linked to the arbitrariness of volume relaxations in charged defect calculations~\cite{bruneval_pressure_2015}. In some cases, for comparison, we nevertheless performed zero pressure calculations of the Raman spectrum (an example is shown in the Supplemental Material~\cite{roma_see_2021}). The OPOs structures have been fully relaxed to a pressure smaller than~1~kbar.
For all configurations the threshold for the relaxation of atomic position has been fixed to 10$^{-3}$ Ryd/Bohr. All DFT and DFPT calculations were performed with the \textsc{Quantum-Espresso} package~\cite{giannozzi_quantum_2009,Giannozzi:2017}.

Extrapolation of the Raman spectrum of point defects in the (B$_{11}$C)C-B-C structure to larger supercells, i.e. smaller defect concentrations, was dealt with using a recently devised procedure~\cite{roma_modeling_2019} that enables us to embed the force constant matrix into that corresponding to a larger bulk supercell. In some cases, for the purpose of comparison with the embedding method,  we performed direct calculations of Raman spectra for 3$\times$3$\times$3 supercells containing more than 400 atoms. For simple point defects the size of the supercell used for the direct calculation (120 atoms) is found to be sufficient, except in cases where the Placzek approximation is not any more justified (see the case of a boron interstitial in charge +1, which we discuss later).

The localisation of phonon modes has been quantified by the variance $\sigma(\nu)$ of the norm of the atomic displacements ($\norm{{\bf u}^i(\nu)}$, i=1,N$_{at}$) of the considered vibrational mode $\nu$:
\begin{equation}
  \sigma(\nu)=\sum_{i=1}^{N_{at}} \left[ \sqrt{{u_x^i(\nu)}^2+{u_y^i(\nu)}^2+{u_z^i(\nu)}^2}-\frac{1}{N_{at}}\sum_{j=1}^{N_{at}}\sqrt{{u_x^j(\nu)}^2+{u_y^j(\nu)}^2+{u_z^j(\nu)}^2}\right]^2;
  \label{EqSigma}
  \end{equation}
this stems from the consideration that the more a vibrational mode is localised, the larger will be the difference between the displacement of the most involved atoms with respect to the less involved ones; we have to keep in mind that phonon eigenvectors are normalized.

Especially when phonon modes are localised we are interested in knowing the character (chain or icosahedral) of vibrational modes. We have defined it by assigning a character to each atom (chain or icosahedron) 
and weighting different atom contributions according to the norm of their displacement. Thus the character of a mode $\nu$ is defined as:
\begin{equation}
  Ch(\nu)=\frac{1}{N}\sum_{j=1}^{N_{at}}ch(j)*\norm{{\bf u}^j(\nu)},
  \label{EqLocalisation}
  \end{equation}
where the sum runs over all $N_{at}$ atoms in the cell and $ch(j)$ and ${\bf u}^j(\nu)$ are, respectively the character and the displacement of atom $j$ within the vibrational mode $\nu$.
The character of each atom is defined through the distance of the neighbors. Defining short bonds as those between 2.6 and 3.0 \AA ~and long bonds as those between 3.0 and 3.5~\AA, we define chain atoms as those having at least one short bond and at most three long bonds and assign them $ch(j)=-1$; icosahedral atoms have at most one short bond and at least 5 long bonds, and have $ch(j)=1$. Atoms not falling in those two categories are attributed an hybrid character, and we assign them $ch(j)=0$ (this is for example the case for boron interstitials).

While plotting calculated spectra we have to assume a certain broadening of Raman lines; experimentally this can come from the resolution of the measuring apparatus as well as from the intrinsic linewidth which depends on the lifetime of the excitations. As the latter broadening is not straightforward to calculate, we simply apply a Lorentzian broadening of width 10 cm~$^{-1}$ to our calculated spectra, which ends up mixing the chain/icosahedral characters of nearby modes; for the sake of visualisation, we enhance the specific character of these mixed modes by a function of the form:
\begin{equation}
  X_\nu=\frac{\alpha Ch(\nu)}{\alpha|Ch(\nu)|+1}
  \label{EqEnhance}
\end{equation}
where $\nu$ is the mode label and $\alpha$ an arbitrary enhancement factor. In the following we use $\alpha=$100 for the 15-atom unit cell and $\alpha/n^3$ for $n\times n\times n$ supercells. Large unit-cells as those of OPOs structures were treated as supercells of equivalent size.

The intensity in the Placzek approximation depends on temperature through the occupation of the phonon modes according to the Bose statistics. It is the only temperature dependence that we consider here, plotting all spectra at T=300 K. In particular, we neglect the effect of thermal expansion on the position of the phonon modes.

Defects formation energies were calculated using the usual approach as in our previous work on vacancies~\cite{gillet_influence_2018}, for charged defects only the monopole Madelung correction was added, to remove spurious interaction between periodic images. The reference carbon chemical potential for carbon-rich conditions is the energy per atom of graphite, while for boron-rich conditions the reference is $\alpha$-boron. 


\subsection{Experimental methods}
Turning to  the experimental procedures, the spectrum in Fig.~\ref{FigExpGrainsFins} was obtained on a sample with a density of 2.42~g/cm$^{3}$, i.e., 96\% of the commonly reported nominal density of B$_4$C of 2.52~g/cm$^3$\cite{domnich_boron_2011}. Grain sizes were between 0.3 and 0.8~$\mu$m; the sample was sintered from HD20 powder.

For the large grain spectra shown in the following (section~\ref{Geometry}), pure B$_4$C pellets (B+C $>$99~\%,
free carbon $<$0.5~weight~\%) with density close to 98~\% of the nominal density were prepared by hot pressing process.
The grain size was measured to be between 10 and 50 $\mu$m.
Raman characterizations were performed, in both cases, with a Renishaw Invia Reflex high-confocal spectrometer equipped
with 633~nm of He-Ne excitation laser and a 1800 groove/mm grating coupled with a Leica microscope (x50).
Raman spectra of large grain samples were carried out for a spectral acquisition between 400 and 1200~cm$^{-1}$. Acquisition times are typically 20s.
The Raman spectrometer was calibrated with silicon single crystals.
Raman spectra were collected every 1~$\mu$m along a line crossing a grain boundary in order to produce a spatial map of the collected intensities. The spatial resolution of the Raman system is 1~$\mu$m.
The standard setup, which was used, collects the spectra in the backscattering geometry.

The excitation laser has a power of 7~mW. We have checked the sample heating by performing several tests at various laser powers (from 0.15 to 7~mW) on the same grain.  For these analysis conditions, neither surface evolution nor sample oxidation are observed. In addition, with similar analysis conditions, we do not observe any defect annealing due to the laser heating in other B$_4$C samples either damaged by ion irradiation or amorphised.

\section{Results and discussion}
\label{Results}

\subsection{Geometry and Raman intensity}
\label{Geometry}
Simulated Raman spectra of crystalline solids imply most frequently averaging over the relative direction of the polarization of the incident and scattered light beams. Such an approach, inherited from Raman spectra of molecules in the gas phase~\cite{porezag_infrared_1996}, is perfectly suited to represent the Raman spectrum of polycrystalline solids, where grains are randomly oriented with respect to the experimental apparatus. In crystals, however, it can be useful to select specific geometries in order to enhance or suppress some specific Raman lines, through the symmetry of the crystal lattice. Recent comparisons of experimental and calculated spectra for silicon carbide polytypes~\cite{roma_linear_2016} provide an example of how to exploit this opportunity. Here we focus on the (B$_{11}$C$^p$)C-B-C structure to show how the geometry of the experimental settings can influence the acquired Raman spectrum.

The geometry of an acquired spectrum is commonly labeled using the Porto notation; to give an example, the notation $Z(YX)\overline{Z}$ means that the light impinges on the sample along the~$Z$ direction and is collected in the backscattering geometry along the same cartesian axis (reversed direction, $\overline{Z}$). The polarizations are indicated in parentheses: the polarization of the incident beam is along the $Y$ axis, while the detector selects the light polarized along $X$ for the scattered beam. Supposing all possible polarizations are analysed for the scattered beam (always perpendicular to the propagation direction of the light beam), then we would have $Z(YX+YY)\overline{Z}$. This is a typical setting for Raman measurements in the backscattering geometry.

\begin{figure}
  \includegraphics[width=\columnwidth]{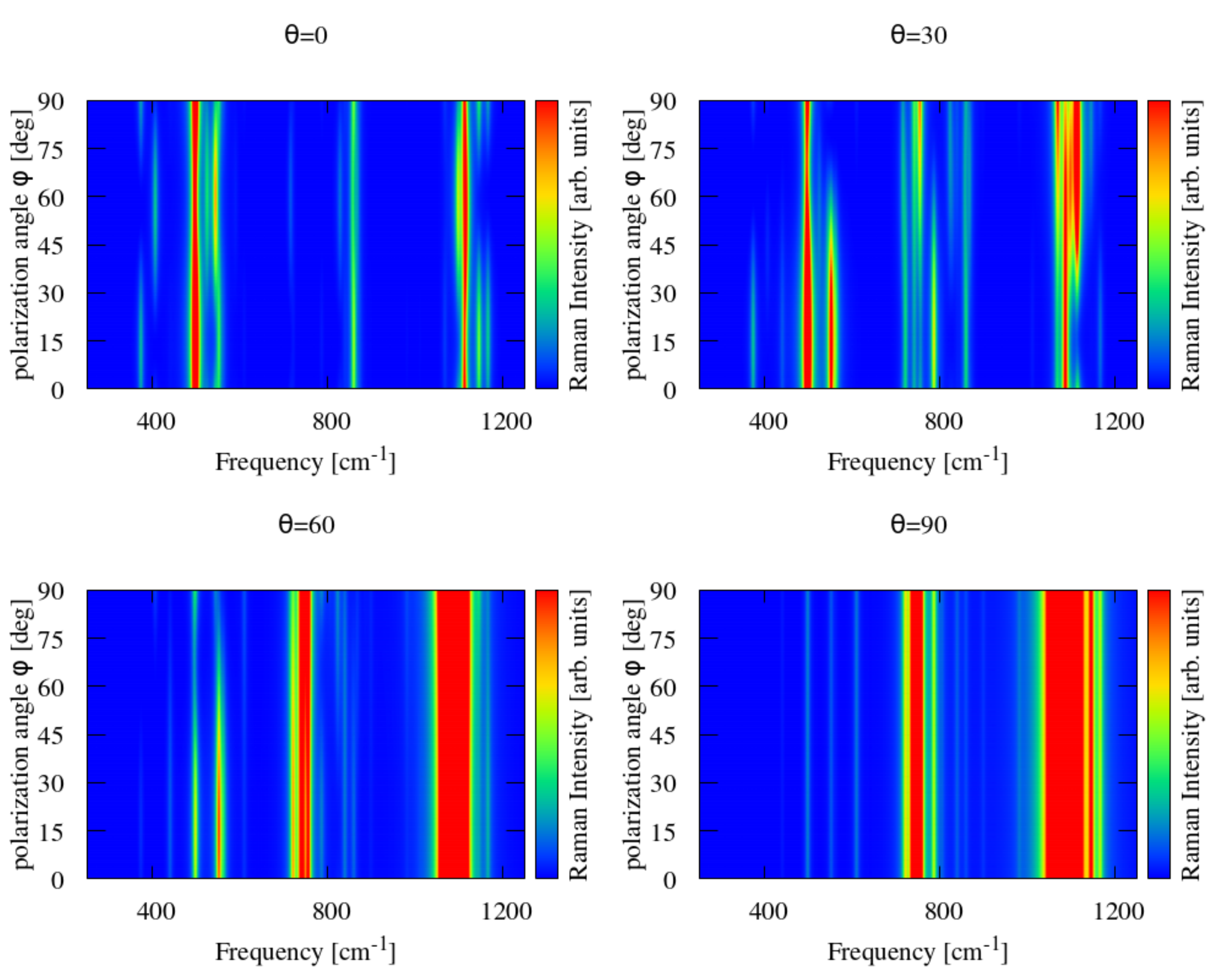}
  \caption{Maps of the computed Raman intensity {\sl vs} polar angle $\phi$ for various choices of the azimuthal angle $\theta$. $\theta$ and $\phi$ indentify the geometrical relationship between the incident light polarization and the crystal lattice orientation. $\theta$ is given in degrees.}
  \label{FigMapsPhi}
\end{figure}

In general, however, the incident beam is not necessarily aligned with a crystallographic direction; this typically happens when doing topography on samples containing large grains whose orientation is not known.
In this case the incident light beam impinges onto the grain along a direction which is identified by two angles, the azimuthal angle $\theta$ {\sl w.r.t.} the $Z$ crystallographic axis and a polar angle $\phi$ between the $X$-axis and the projection of the incident direction on the $XY$ crystallographic plane; for the boron carbide structure we fix the $Z$ axis as the $\langle 111\rangle$ direction of the trigonal reference system and the cartesian $XY$ plane coincides with the $\langle 111\rangle$ plane of the non-cartesian crystal reference system.
For given $\theta$ and $\phi$ we can predict the Raman spectrum in the backscattering geometry, which in general corresponds to an arrangement $A_{\theta,\phi}(BC+CC)\overline{A}_{\theta,\phi}$, where directions B and C are perpendicular to  A and to each other.
$\phi$ maps of calculated Raman spectra obtained for various possible values of the azimuthal angle $\theta$ are shown in Fig.~\ref{FigMapsPhi}.

\begin{figure}
  \includegraphics[width=\columnwidth]{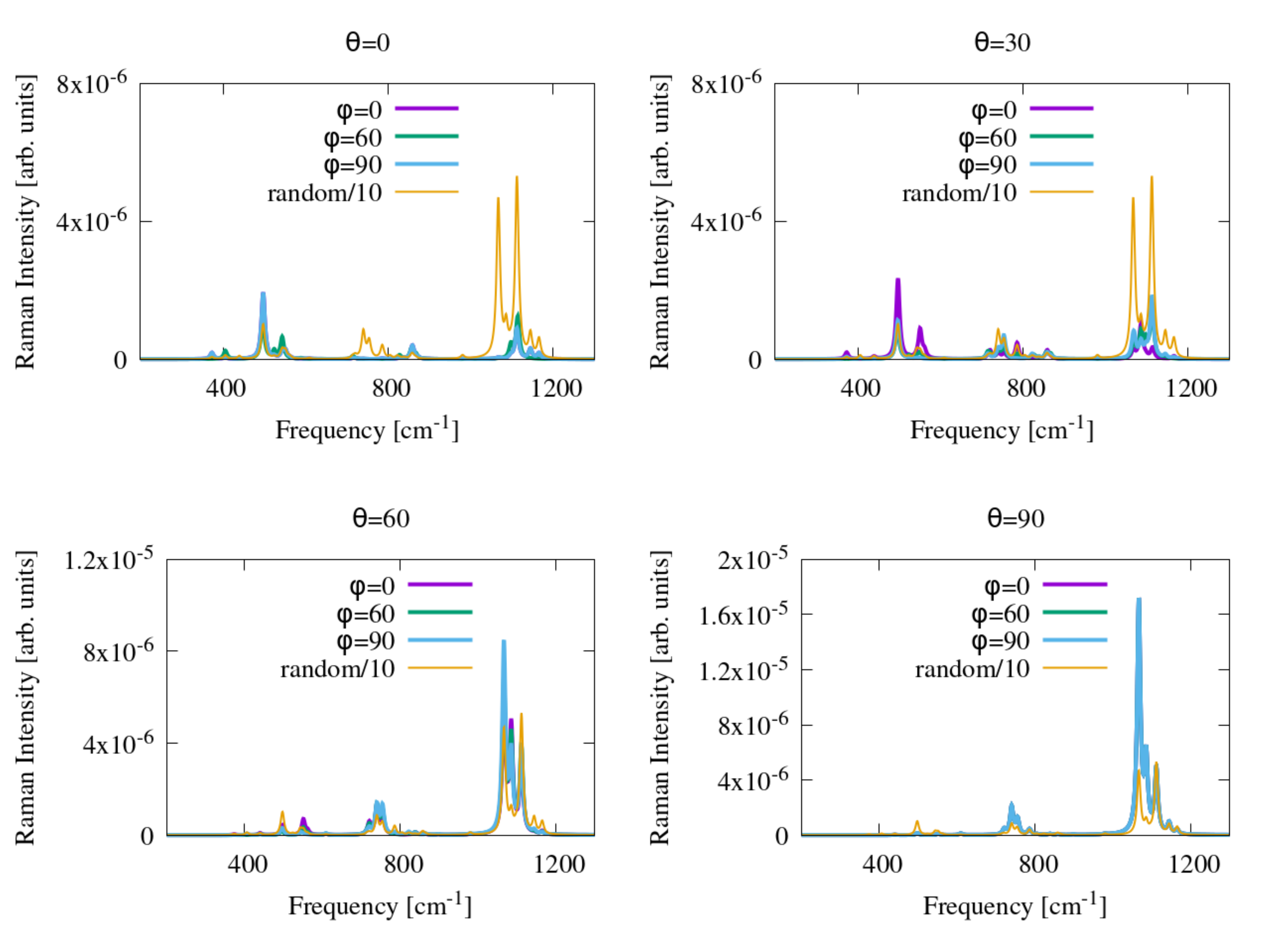}
  \caption{Comparison of simulated spectra for selected values of the azimuthal ($\theta$) and polar ($\phi$) angles. The spectrum corresponding to a polycrystalline sample with randomly oriented grains, shown here for comparison (yellow curves), is reduced by a factor of ten.}
  \label{FigGeomRand}
\end{figure}

From these maps it is clear that the choice of the incident light beam direction with respect to the crystallographic axis has a crucial influence on the presence and intensity of several features of the spectrum.
To further stress this points we show spectra simulated for a few azimuthal angles ($\theta$=0$^\circ$, 30$^\circ$, 60$^\circ$, 90$^\circ$) and a few choices of the polar angle $\phi$ in Fig.~\ref{FigGeomRand}. Here we show for comparison also the calculated spectrum that we expect for a fully random orientation of the grains.


Figures \ref{FigMapsPhi} and \ref{FigGeomRand} predict the modifications occuring to the Raman spectrum  when modifying the geometrical settings of an experiment on a single crystal of (B$_{11}$C)C-B-C, and how different they are with respect to a random spectrum, i.e., a spectrum obtained on powder or polycrystalline sample. Let us consider now collecting a Raman spectrum on a sample whose grains size is comparable to the measuring spot : the outcome will not necessarily look like a random spectrum; clearly, if the spectrum is acquired on a few large grains and, in particular, if it is acquired on one single grain, the outcome will depend on the grain orientation {\sl w.r.t.} the propagation direction of the incident light beam. In particular the intensity of the two lines slightly below and slightly above 500 cm$^{-1}$, which are attributed respectively to the rotation of the chain and to the libration of the icosahedra~\cite{jay_conception_2016, vast_atomic_2000}, is gradually suppressed with the increase of the azimuthal angle. Moreover, the strong lines around 1100 cm$^{-1}$, which are attributed to two $A_{1g}$ modes of symmetric streching, respectively the stretching of the intericosahedral bonds and the streching of the chain, decrease in intensity when the light beam impinges along the $\langle 111\rangle$ direction of the trigonal crystal structure ($\theta=0$).

The discussion of the previous paragraph helps us to understand the experimental results which we have obtained on a sample having large grains, whose size (several microns) is much larger than the laser spot used for the Raman measurement.

\begin{figure}
  \includegraphics[width=\columnwidth]{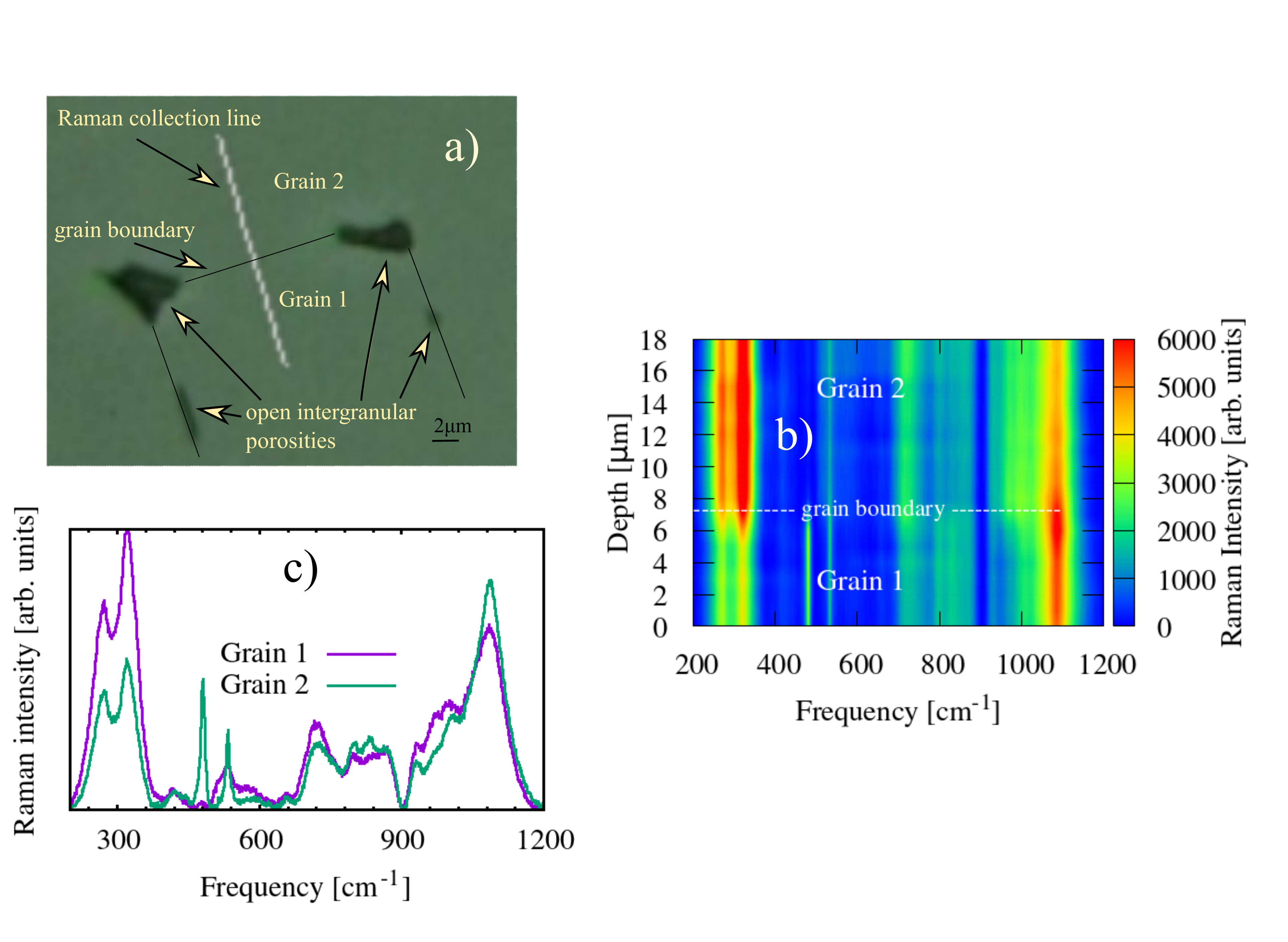}
\caption{Experimental Raman spectra of boron carbide across a grain boundary. a) Micrograph showing the Raman collection line and the grain boundary. b) Spatial map of the spectrum along the line shown in panel~a). c) Comparison of the spectra of grains 1 and 2. }
  \label{FigGB12}
\end{figure}

In Fig.~\ref{FigGB12}b we show a spatially resolved experimental Raman spectrum clearly showing relevant modifications occuring when crossing the grain boundary (see the Raman collection path in Fig.~\ref{FigGB12}a). In particular, the first peak, at~480 cm$^{-1}$, attributed to the rotation of the chains ~\cite{jay_conception_2016,vast_atomic_2000},   
disappears when passing from grain 1 to grain 2 (see panels b and c of Fig.~\ref{FigGB12}). Other less striking, though visible, modifications of the spectrum occur: slightly more pronounced contributions in the region 950-1000 cm$^{-1}$, as well as at 700 cm$^{-1}$, attributed to two icosahedral vibration modes~\cite{jay_conception_2016}.
The identification of the grain boundary, although not striking from the micrograph in Fig.~\ref{FigGB12}a, was clearly observed during acquisition and is confirmed by the shape of the open intergranular porosities (dark regions): the two largest ones are identified as triple points, which suggests the location of the grain boundary connecting them. Further hints can be obtained from the Raman spatial maps in the Supplemental Material~\cite{roma_see_2021}, Figure S4.

The present analysis may shed some light on a longstanding controversy on the Raman spectrum of boron carbide~\cite{kuhlmann_improved_1993} which was recently revived~\cite{werheit_systematic_2019}. According to Refs.~\onlinecite{kuhlmann_improved_1993} and \onlinecite{werheit_systematic_2019} the usually accepted Raman spectrum of boron carbide~\cite{tallant_boron_1989}, which is also mostly in agreement with theoretical spectra for the (B$_{11}$C)C-B-C structure~\cite{jay_carbon-rich_2014,jay_conception_2016} and with our polarization averaged spectrum, would stem essentially from boron carbide surface and not from the bulk, due to absorption of commonly used laser frequencies in the first 100 nm or so. It is true that, although a clear understanding of the band gap of boron carbide is still a matter of active study~\cite{Shirai:2014,rasim_local_2018,ektarawong_structural_2018}, laser wavelenghts of 514~nm or 532~nm correspond to energies slightly higher than measured values of the band gap; however, for another carbide with similar band gap (cubic silicon carbide), in spite of some baseline due to fluorescence, the measured spectra can definitely be identified with the bulk spectrum~\cite{roma_linear_2016,miro_monitoring_2014}. A more detailed study of the influence of the incident laser wavelength~\cite{xie_microstructural_2017} shows that, although modifications occur, in particular concerning the relative intensity of the main peak at 1090 cm$^{-1}$ and the low frequency features around 300 cm $^{-1}$, the main features of the spectrum are maintained, including the doublet at 480/535 cm$^{-1}$, respectively attributed to the pseudo-rotation of the chains and to the libration of the icosahedra~\cite{jay_conception_2016}; the latter is, in contrast, virtually absent from the spectrum labeled ``bulk spectrum'' in Ref.~\onlinecite{werheit_systematic_2019}.
Concerning this doublet, we showed here that it is well present even with a subbandgap exciting frequency, as in~\cite{xie_microstructural_2017}, and we predict how it disappears in specific grain orientations, as already remarked qualitatively from experiment~\cite{tallant_boron_1989}. Even the intensity of the main  peaks around 1100 cm$^{-1}$  varies strikingly according to the grain orientation, so that, unless a large enough number of grains is probed, the acquired spectrum is clearly not an averaged one.

\subsection{Mode localisation and character}
\label{Localisation}
There is another point we shall discuss about the Raman spectrum of the (B$_{11}$C)C-B-C structure before coping with the influence of point defects: it is the question of mode localisation and character. In order to link the features of a measured Raman spectra to specific atomic modifications it is customary to identify some signals in the spectrum as icosahedral or chain vibrations. For example, the region of the spectrum above 600~cm$^{-1}$, and in particular the main peak at $\sim$~1080~cm$^{-1}$, has been attributed to vibrations in the icosahedral units~\cite{tallant_boron_1989,xie_microstructural_2017} or to the symmetric chain strectching~\cite{jay_conception_2016}, while the doublet around 500 cm$^{-1}$ is  associated to the chain rotation, for the lowest peak, and to the icosahedral libration for the highest one~\cite{lazzari_atomic_1999,vast_atomic_2000}. From theoretical calculation of Raman spectra, one can look at the displacement patterns of the modes having the largest Raman cross section and check which are the atoms with the largest displacements~\cite{jay_conception_2016}. Although a qualitative inspection of displacement patterns can already provide useful information on the chain or icosahedral character of the mode, it is not easily applicable to large supercells containing defects. Therefore, in the present work, we use the quantitative approach described in section \ref{Method}. Equation \ref{EqLocalisation} allows us to assess the character of specific features of the spectrum, according to the contribution of all the modes contributing to a Raman peak, and with the presence of defects.
Moreover, in particular when coping with point defects, we also would like to assess whether modes responsible of (or contributing to) specific Raman peaks are more or less localised. We have thus used the variance (equation \ref{EqSigma}) of the atomic displacements (see section \ref{Method}) to quantify mode localisation. 
\begin{figure}
  \includegraphics[width=0.5\columnwidth]{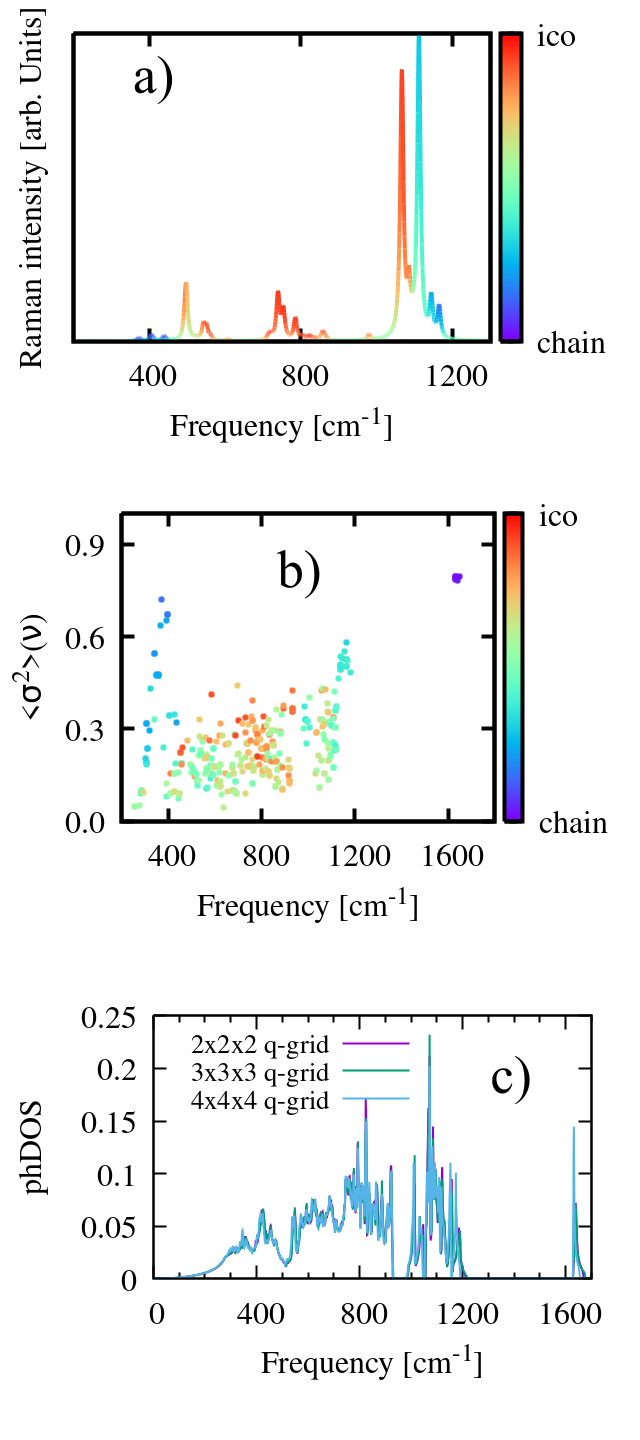}
  \caption{Theoretical bulk spectrum of carbon-rich boron carbide, B$_4$C in the 15-atom ground state structure (B$_{11}$C$^p$)C-B-C. a) the calculated first order Raman spectrum (broadening: 10~cm$^{-1}$) with chain/icosahedron character enhanced by a factor $\alpha=100$ following equation \ref{EqEnhance}. b) the localisation of each mode (irrespective of its Raman activity) given by equation \ref{EqLocalisation} as a function of frequency (all modes correponding to a 2$\times$2$\times$2 supercell are included). c) The normalized phonon density of states, obtained with three different {\bf q}-points grids in the BZ, for comparison. The infrared active mode at 1641~cm$^{-1}$ is the antisymmetric chain stretching. Its Raman intensity is approx. 1/100 of the intensity of the largest peak and for this reason it is not shown on panel a).}
  \label{FigBulkLoc}
\end{figure}

In panel~a) of Fig.~\ref{FigBulkLoc} we show the calculated random spectrum of the (B$_{11}$C)C-B-C structure where the color of the line indicates the character (chain/icosahedral) of the involved modes. In Fig.~\ref{FigBulkLoc}b we show the localisation of all modes (including those that are not Raman active) as a function of frequency. Their character is indicated by their color, with the same scale as in Fig.~\ref{FigBulkLoc}a. We see that, although well localised modes with chain character are clearly present below 400 cm$^{-1}$, they are not Raman active (see also the comparison with the phonon density of states in Fig.~\ref{FigBulkLoc}c). The lowest peak of the doublet around 500~cm$^{-1}$, has indeed a character which is less clearly icosahedral than the highest one, but still not clearly identified as a chain mode with the present analysis tool, while a direct analysis enables us to attribute it to the chain rotation~\cite{jay_conception_2016}. Conversely, the second of the two main peaks ---unresolved by experiment--- above 1000~cm$^{-1}$ has a clear chain character, while the first peak is even more clearly an icosahedral mode. The mode characters are in agreement with those inferred from the analysis of the calculated eigenvectors of the modes with the highest Raman activity~\cite{jay_conception_2016}.


\subsection{Point defects and their influence on the Raman spectrum}
\label{Defects}

In this section we present our results on the influence of point defects on the Raman spectrum of boron carbide. Some theoretical studies devoted to predicting how alternative structures may modify the Raman spectrum of the reference (B$_{11}$C)C-B-C have been performed~\cite{jay_carbon-rich_2014,jay_conception_2016,kunka_evaluating_2017}, however, there are two aspects that were up to now overlooked: the point defect character of such structural modification and their actual concentration. Concerning this last point, although point defects can be present in concentrations larger than equilibrium ones, due, e.g., to the kinetics of material synthesis, it seems reasonable to consider first those defects whose equilibrium concentration is the highest one. To approach this issue from a point defect perspective we first give an introductory note on point defects in boron carbide.

\subsubsection{Stability of simple defects in (B$_{11}$C)C-B-C}
\label{DefectsFormationEnergies}
It is generally accepted that defects are unavoidable in boron carbide and their kind and distribution modifies, and at least partially controls, the stoichiometric ratio between boron and carbon after a high temperature synthesis~\cite{balakrishnarajan_structure_2007,domnich_boron_2011,werheit_advanced_2012}. Several works, with a first principles approach or empirical potentials, investigated the energetics of several structural variations of boron carbide~\cite{saal_structural_2007,raucoules_mechanical_2011,betranhandy_ab_2012,wang_predicted_2014,ektarawong_configurational_2015,yao_phase_2017,schneider_stability_2017,ektarawong_structural_2018,gillet_influence_2018, rasim_local_2018}. Most of them coped with modifications in the 15-atom unit cell, other investigated larger structures, and thus more complex arrangements and/or lower defect concentrations; however, apart from our previous study of the negatively charged vacancy~\cite{jay_conception_2016} and our recent complete work on vacancies~\cite{gillet_influence_2018}, all of them are limited to charge neutral defects. To the best of our knowledge no systematic theoretical study of the stability of point defects and their possible charge states in semiconducting boron carbide exists.

\begin{figure}
  \includegraphics[width=0.6\columnwidth]{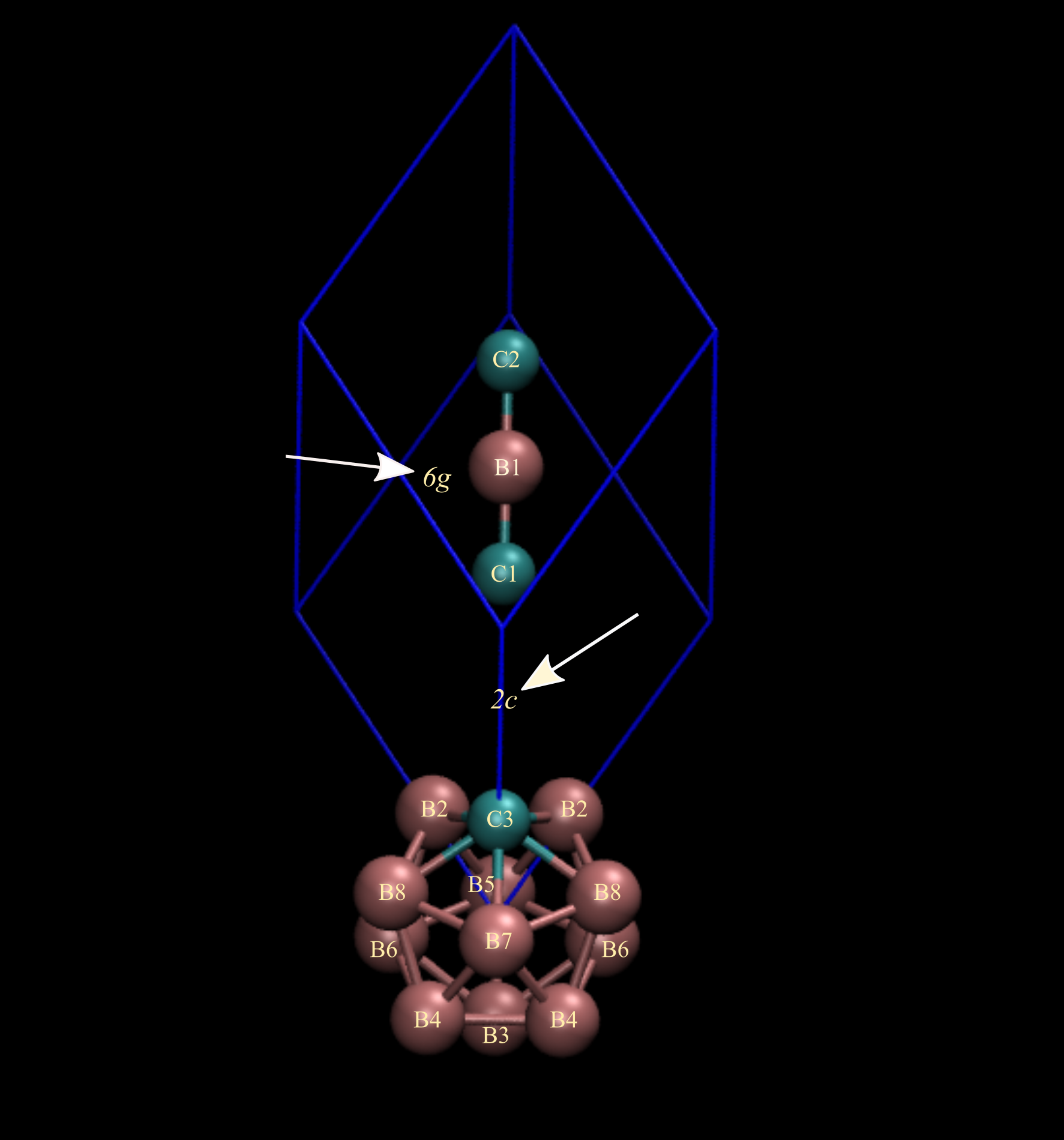}
  \caption{The labelling of the atomic sites of the  15-atom (B$_{11}$C)C-B-C unit cell (in blue). The two arrows indicate the approximate position of the 2$c$ and 6$g$ EIWS (extra interstitial Wyckoff sites) of the $R\overline{3}m$ rhombohedral space group. Among icosahedral atoms, equatorial borons B5-B8 are bound to neighboring C-B-C chains, while polar atoms B2-B4 and C3 bind to neighboring icosahedra. }
  \label{FigSiteNames}
  \end{figure}

We performed recently such a systematic study \cite{gillet_influence_2018,roma_see_2021} and,
although the main goal of this paper is to discuss the Raman signature of defects, we need to summarize here our results on the stability and charge states of point defects, as they have not yet been published in their entirety. The main results are summarized in Table~\ref{TabAllDefects}. Further details are given in the Supplemental Material~\cite{roma_see_2021}.
Simple point defects in B$_4$C boron carbide can be divided in three categories: vacancies, interstitials, and antisites.
The labelling of sites of the 15-atom (B$_{11}$C)C-B-C unit cell are shown in figure~\ref{FigSiteNames}

\begin{table}
  \caption{Simple defects and a few defect complexes in B$_4$C boron carbide, with the calculated fixed volume formation energies and Fermi level at the valence band top, $\mu_e$=0), for a few charge states (-1/+1 means one added/removed electron); in this column $m$ indicates an odd number of electrons (metallic configuration). The last column gives, when applicable, the figure number where the corresponding calculated Raman spectrum is shown. EIWS stands for ``Extra-Interstitial Wickoff Site''. See Figure~\ref{FigSiteNames} for the labelling of atomic sites in the 15-atom unit cell.}
  \begin{tabular}{l c c c c c}\hline\hline
    Defect type & structure &        E$_f^{C-rich}$($\mu_e$=0) & E$_f^{B-rich}$($\mu_e$=0) & charge state & Raman Fig. \# \\
                &           &           [eV]                  &  [eV]                    &              &               \\  \hline\hline
vacancies &   boron chain V$_{B1}$ (C-$\square$-C)      & 1.62 &  1.77   & 0 $m$       & -- \\
 &    boron chain V$_{B1}$ (C-C) & 2.16    & 2.31 & -1       & \ref{FigVacancies}a \\
 &   carbon chain V$_{C2}$     & 2.81 &  2.66  & 0        & -- \\
 &   carbon chain V$_{C1}$     & 2.96 &  2.81  & 0        & -- \\ 
 &   carbon chain V$_{C1}$     & 3.82 &  3.67  & -1 $m$      & -- \\ 
 &   carbon chain V$_{C2}$     & 3.94 & 3.79   & -1 $m$      & --  \\ \hline
antisites &
    carbon icosahedral C$_{B4}$ & 0.04  & 0.77 & +1       & \ref{FigAntisites}a\\
  &  carbon icosahedral C$_{B5}$ & 0.28 & 1.02  & +1       & --\\
  &  carbon icosahedral C$_{B2}$ & 0.31 & 0.43  & +1       & --\\
  &  carbon icosahedral C$_{B6}$ & 0.41 & 1.14  & +1       & --\\
  &  carbon icosahedral C$_{B7}$ & 0.82 & 1.55  & +1       & --\\
  &  carbon icosahedral C$_{B3}$ & 0.84 & 1.57  & +1       & \ref{FigAntisites}c\\
  &  carbon icosahedral C$_{B1}$ & 0.94 & 1.68  & +1       & --\\
  &  carbon icosahedral C$_{B8}$ & 0.95 & 1.68  & +1       & --\\
  &  boron icosahedral B$_{C3}$ ($\equiv$ B$_{12}$) & 1.24 & 0.52  & -1 & \ref{FigAntisites}e \\ \hline
 boron interstitials &
    EIWS 2c, icos. B side  I$_B$ & 2.36 & 2.21 &+3      & \ref{FigInterstitials}c \\
  &  EIWS 2c, icos. B side I$_B$ & 4.39 & 4.24 & +1      & \ref{FigInterstitials}a \\
  &  EIWS 2c, icos. C side I$_B^*$ & 2.90 & 2.75 & +3     & -- \\
  &  EIWS 2c, icos. C side I$_B^*$ & 3.26 & 3.11 & +1     & -- \\
  &  EIWS 6g, C$\langle^\text{B}_\text{B}\rangle$C & 2.60 & 2.45  & +3 & \ref{FigInterstitials}g \\ 
  &  EIWS 6g, C$\langle^\text{B}_\text{B}\rangle$C & 3.73 & 3.59 & +1 & \ref{FigInterstitials}e \\ \hline
    %
 carbon interstitials &  EIWS 6g, C$\langle^\text{C}_\text{B}\rangle$C & 2.82 & 3.40  & +2 & -- \\ 
  &  EIWS 2c, icos. B side I$_C^*$ & 4.84 & 5.43 & +2 $m$    & -- \\
  &  EIWS 2c, icos. C side I$_C^*$ & 4.83 & 5.41  & +1     & -- \\ \hline
 defect complexes &   bipolar defect: C$_{B3}^{+}$+B$_{C3}^{-}$   & 0.24 & 0.24 & 0 & \ref{FigBipolar}a \\
  &  V$_{B1}^-$+C$_{B4}^+$ complex & 1.48 & 2.36 & 0 & -- \\
  &  B$\langle^\text{B}_\text{B}\rangle$B chain (or B$_{C1}$+B$_{C2}$+I$_B$) & 4.70 & 3.00   & +1 & \ref{FigBBBB}a \\
  &  B$\langle^\text{B}_\text{B}\rangle$B chain (or B$_{C1}$+B$_{C2}$+I$_B$) & 3.89 & 2.19   & +3 & \ref{FigBBBB}c \\
  &  3+1 defect: I$_B^{3+}$+3B$_{C3}^-$ & 2.39 & -0.07 & 0  & \ref{Fig3+1} \\ \hline\hline
  \end{tabular}
  \label{TabAllDefects}
\end{table}

Concerning vacancies, for which we published a complete report recently~\cite{gillet_influence_2018}, we know that the boron chain vacancy is by far the more stable one~\cite{betranhandy-2012,raucoules_mechanical_2011,jay_conception_2016,schneider_stability_2017,you_first-principles_2018,gillet_influence_2018}, with a formation energy smaller than~2~eV, while the formation energy of carbon chain-end vacancies is around 3~eV. Icosahedral vacancies not only have a much higher formation energy, but they are very unstable, because the barrier to convert them into chain vacancies is very low or vanishing, which means that, even under irradiation, they are transient species, thus virtually absent. Chain vacancies are neutral in $p$-type boron carbide and negatively charged when the material is intrinsic or $n$-type~\cite{gillet_influence_2018}. The charge transition level is approximately 0.5~eV above the valence band for the boron chain vacancy and 1~eV for carbon ones. A remarkable fact is that the structure of the boron chain vacancy changes upon electron capture, forming a C-C bond~\cite{jay_conception_2016,gillet_influence_2018}, similar to what happens under pressure~\cite{raucoules_mechanical_2011}.  

Antisites have been widely studied in the framework both of chain variants~\cite{saal_structural_2007,raucoules_mechanical_2011,betranhandy_ab_2012} and also including icosahedral substitutions~\cite{ektarawong_configurational_2015,yao_phase_2017,ektarawong_structural_2018}, but always in the neutral state. 
Our results show, conversely, that carbon antisites are almost always in a positive charge state, while boron ones are mostly negative. The most stable antisite, formed when a carbon atom substitutes a polar boron atom on the opposite side (but not antipodal) with respect to the icosahedral carbon, is a negative-U defect, with a +1/-1 charge transition level close to the bottom of the conduction band; its formation energy almost vanishes under extreme $p$ doping. This antisite is named in the following C$_{B4}$ while the antipodal one is called C$_{B3}$ (see page 2 of the Supplemental Material~\cite{roma_see_2021} for details).

The most stable boron antisite is, as could be expected, the icosahedral one; it is neutral only in strongly $p$-type boron carbide, otherwise it is negatively charged. Its formation energy vanishes when the Fermi level is smaller than 1.5~eV above the valence band top.

Interstitial defects have not been considered on their own, up to now, in boron carbide, to the best of our knowledge, but they have been studied as components of large crystalline motifs~\cite{rasim_local_2018,ektarawong_structural_2018} recently shown to provide the ground states for boron-rich boron carbides~\cite{jay_theoretical_2019}. Such structures, which eventually explained the semiconducting character of boron-rich boron carbide, show, when averaged in a 15-atom rhombohedral structure, ordered partial occupation (OPO) of some specific interstitial Wyckoff crystallographic sites~\cite{jay_theoretical_2019} and can be seen as an ordered  combination of boron interstitials ---of at least two kinds--- and boron antisites. Our study of interstitials in the (B$_{11}$C)C-B-C structures show that both boron and carbon interstitials can assume a variety of charge states going from +3 to -2. They have relatively high formation energies; the most favorable cases are boron interstitials in +3 charge state in strong $p$ conditions, where their formation energy is between 2 and 3~eV. Interstitial boron occurs either at the 2$c$ or at the 6$g$ Wyckoff sites of the rhombohedral $R\overline{3}m$ space group. The former, which has two variants ---either close to the carbon side of the icosahedron or to the boron side (see Table ~\ref{TabAllDefects})---, is labeled I$_B$ in this work. The latter, which relaxes in a pantograph-like structure, is called C$\langle^\text{B}_\text{B}\rangle$C (for the structures see figure S1 in the Supplemental Material~\cite{roma_see_2021} and also Table II of Ref.~\onlinecite{jay_theoretical_2019}). Similarly, B$_4$ blocks occur in a pantograph structure in the mentioned OPO$_1$ structure~\cite{jay_theoretical_2019} and are thus referred to as  B$\langle^\text{B}_\text{B}\rangle$B (see Table~\ref{TabAllDefects}).

The results show that boron and carbon antisites are the defects with lowest formation energies and, given their charge states, they induce $p$ self-doping in B$_4$C, pinning the Fermi level close to the valence band top. More details on the results on point defects that we have just summarized are in the Supplemental Material~\cite{roma_see_2021}.

\subsubsection{From charge-compensated defect complexes to complex (meta)stable boron-rich phases}
\label{DefectComplexes}
Concerning possible association of defects, at least one low formation energy defect complex has already been studied, the bipolar defect~\cite{mauri_atomic_2001,raucoules_mechanical_2011}. It is formed by a (B$_{12}$) icosahedron plus a (B$_{10}$C$_2$) one. This stoichiometric defect has a formation energy as low as 0.24~eV in the neutral state; we highlight here the possible role of charge compensation, because we can consider it, in the light of our results, as formed by a positive carbon antisite and a negative boron one, each on a different icosahedron. Seen as such, in the light of our calculated formation energies for the two isolated defects, we can conclude that the bipolar defect, which is an antisite pair, is stabilized by a strong Coulomb binding energy between the two antisite defects.

In the same spirit we have calculated the formation energy of a few other neutral defect associations. One is the complex formed by a negative vacancy and a positive carbon antisite ---a (B$_{10}$C$_2$) icosahedron---; the formation energy is definitely higher than the bipolar defect (1.5~eV in C-rich conditions) but still with a sizeable binding energy.

Developing further the idea of charge compensation, and by analogy with the OPOs structures, we considered a defect complex which turned out to be the ground state structure at low temperature at~19.2~\% atomic carbon concentration within DFT-LDA~\cite{Note_LDA-GGA_3+1}. It is formed by the combination of a boron interstitial in the 2$c$ Wyckoff position of the rhombohedral 15-atom unit cell, and three boron icosahedral antisites (i.e., three B$_{ 12}$ icosahedra) surrounding it; such a defect has a very low formation energy within DFT-LDA, placing the structure with one defect in a 3$\times$3$\times$3 (B$_{11}$C)C-B-C supercell (a 406-atom motif) on the convex hull at 19.2~\% carbon concentration and thus enhancing  the known convex hull of boron carbide structures~\cite{jay_theoretical_2019}, in the limit of low temperature. The formation enthalpy of this structure is reported in Table~S1 of the Supplemental Material~\cite{roma_see_2021} (see also figure S2 for the convex hull). We dub the defect complex as ``3+1'' (Table~\ref{TabAllDefects}) and the new phase as ``(3+1)@27'', in reference to the three antisites and one interstitial out of 27 unit-cells. 

The ``(3+1)@27'' phase can be built by embedding simple defects of the (B$_{11}$C)C-B-C structure into a large crystal motif of 406 atoms, a size similar to that of that of the OPOs structures (414 atoms). A comparison of the energetics and stoichiometry of these phases can be found in Table S1 of the Supplemental Material~\cite{roma_see_2021}.
The large motifs of these perfectly ordered ground state phases is
caused by the need of charge compensation. 
Analogues could be the so-called ordered vacancy compounds (OVC) of the copper poor region of the phase diagram of ternary semiconductors like CuInSe$_2$ or CuInS$_2$ or their alloys~\cite{zhang_defect_1998}. In that case, the Coulomb interaction between periodically arranged defect complexes formed by two copper vacancies and an indium antisite lowers the energy giving phases like CuIn$_5$Se$_8$. Here the OPOs are formed by a peculiar arrangement of boron interstitials of the 15-atom rhombohedral unit-cell  (plus a few other chain variants) associated to a careful packing of B$_{12}^-$ icosahedra that enables charge compensation. In the case of OPO$_2$ it is easy to identify the charge compensation: for each boron interstitial (in the 15-atom description) I$_B^{+3}$, the structure has 3 B$_{12}^-$ icosahedra.
The same kind of compensation mechanism occurs for the ``(3+1)@27'' phase, where, however, only three icosahedra are B$_{12}^-$ out of 27, the others being B$_{11}$C.

\subsubsection{Complex motif stable phases from a point defect viewpoint}
\label{Defects2Phases}
It is precisely playing with the idea of associating defects in such a way to provide local charge compensation that we devised the ''3+1'' structure and, probably, further phases with large unit cells could be found in the stoichiometry range between the '``(3+1)@27'' and more boron-rich phases like the OPOs.

The ``3+1'' defect complex is thus formed by four elementary defects, one interstitial and three antisite atoms. This defect, at concentration higher than in the ``(3+1)@27'' phase (in the 2$\times$2$\times$2 supercell of the 15-atom unit cell, with three B$_{12}^-$ out of eight icosahedra, or ``(3+1)@8''), gives a structure which is only slightly above the convex hull (See Supplemental Material~\cite{roma_see_2021}, Tables S1 and S2, and Figure S2).

For this reason, although the OPOs structures are very different from the ``(3+1)@27'' phase, in that they contain many more boron interstitials and not a single B$_{11}$C icosahedron, we dare look at them as combinations of boron interstitials and antisites of the (B$_{11}$C)C-B-C 15-atom unit cell. For example the OPO$_2$ would consist of B$_4$C with 9 boron interstitials and 27 icosahedral boron antisites, for a total of 36 point defects. For the OPO$_1$ we can also count the number of defects; we consider each ~B$\langle^\text{B}_\text{B}\rangle$B block as constitued by two chain boron antisites and a boron interstitial. Then the OPO$_1$ phase, whose icosahedra are all B$_{12}$, has one third of B$_4$ blocks, which means having 27~icosahedral boron antisites, 18~boron chain antisites and 9~boron interstitials for a total of 54~defects. In both OPOs structures, the crystalline motif contains 414 atoms.  A structure with such a large number of point defects might seem to defy the very definition of point defects;  it is nevertheless tempting to remark that the formation energy {\sl per defect} is remarkably very similar in the OPOs and the ''3+1'' structures (0.4-0.6 eV per elementary defect, see the Supplemental Material for further analysis~\cite{roma_see_2021}, Table S2). 

The same holds also for a variant of the OPO$_2$ structure, where boron interstitials atoms, instead of occupying 2$c$ Wyckoff positions, are inserted near to the chain boron atom (Wyckoff position 6$g$ of the rhombohedral unit-cell), forming thus a ~C$\langle^\text{B}_\text{B}\rangle$C pantograph-like structure. Within the LDA approximation this structure (OPO$_2^{panto}$) is slightly more stable than the reported OPO$_2$ structure, at variance with gradient corrected functionals (see the Supplemental Material~\cite{roma_see_2021} for a comparison LDA-GGA, Table S1 and Figure S2).

To further elaborate on charge compensation, now in general and not between charged point defects, the B$_4$C and OPO$_2$ (or OPO$_2^{panto}$) phases (with respectively 15 and 414 atoms per unit-cell) are stable because the extra electrons in the chains compensate the electron-deficient icosahedra by filling the bands that correspond to bonding orbitals:
one C-B-C chain compensate one B$_{11}$C icosahedron in B$_4$C, and two C-B-C plus one C-B-C$\cdots$B (C$\langle^\text{B}_\text{B}\rangle$C) chains compensate three B$_{12}$ icosahedra for OPO$_2$ (OPO$_2^{panto}$)~\cite{jay_theoretical_2019}. The additional electrons of C-B-C$\cdots$B chains can complete the filling of bonding orbitals. Note that the OPO$_2$ and OPO$_2^{panto}$ unit-cells have to be large enough (27 icosahedra) to avoid neighboring chain variants (see Fig.~9 of Ref.~\onlinecite{jay_theoretical_2019}).

If we compare the band structures of (B$_{11}$C)C-B-C and (B$_{12}$)=C-B-C (both with 15 atoms unit cells) we see that they are almost superposables~\cite{kleinman_1991}, with the difference that the last occupied band of (B$_{12}$)C-B-C is half-filled, making it metallic. In the OPO$_2$ structure, as in the `''(3+1)@27'', each C-B-C$\cdots$B blocks provides three additional electrons, which can fill the half filled top of the valence band for each of the three B$_{12}$ icosahedra and recover the semiconducting character of the material.


\subsubsection{Raman spectra of simple defects in (B$_{11}$C)C-B-C}
\label{RamanSimple}
In our study of the Raman spectrum we consider defects whose formation energy is sufficiently low or defects that can be identified as building blocks of the stable phases discussed in section~\ref{DefectComplexes}.
We namely calculated the Raman spectrum of the following single defects: the negative chain-boron vacancy, the boron interstitial in 2$c$ Wyckoff position (I$_B$, in charge states +1 and +3), the boron interstitial forming a sort of pantograph ~C$\langle^\text{B}_\text{B}\rangle$C structure (by analogy with the ~B$\langle^\text{B}_\text{B}\rangle$B block for OPO$_1$ in~\cite{jay_theoretical_2019}), in charge states +1 and +3, two icosahedral carbon antisites in charge state +1 ---(B$_{10}$C$_2$) in antipodal and quasi-antipodal configurations, here dubbed C$_{B3}$ and C$_{B4}$---, the icosahedral boron antisite in negative charge state ---(B$_{12}$) icosahedron---; we also consider the neutral bipolar defect, which is in fact an antisite pair (a boron antisite, giving a B$_{12}$ icosahedron, and a carbon one, a B$_{10}$C$_2$ icosahedron), the ~B$\langle^\text{B}_\text{B}\rangle$B blocks (which can be seen as two chain boron antisites plus a boron interstitial, see Table~S2~\cite{roma_see_2021}), and the ``3+1'' defect.

We investigated a limited number of charge states according to two guidelines: the first is to cope with defects that, due to their formation energies, are more likely to be present, especially in $p$-type boron carbide; the second guideline is, in fact, a constraint, because first order, non-resonant, Raman spectrum is only meaningful for a system with a band gap, which means that the last occupied Kohn-Sham orbital should have an even number of electrons. We neglect here the case in which a sufficiently large spin splitting would produce an insulating configuration with an odd number of electrons, for which, by the way, the calculation of the Raman spectrum would not be  straightforward. Furthermore, according to our calculated charge transition levels (see Supplemental Material~\cite{roma_see_2021}), for most defects, odd number of electron configurations are excluded by the negative-U behavior, or are stable only within a very narrow Fermi-level range; some of them are shown in Table~\ref{TabAllDefects} and labeled by $m$, for {\sl metallic}, in the charge state column.
For the sake of illustration we nevertheless consider, for interstitial defects, both +1 and +3 charge states, although singly charged interstitials have relatively high formation energies.

\begin{figure}
  \includegraphics[width=\columnwidth]{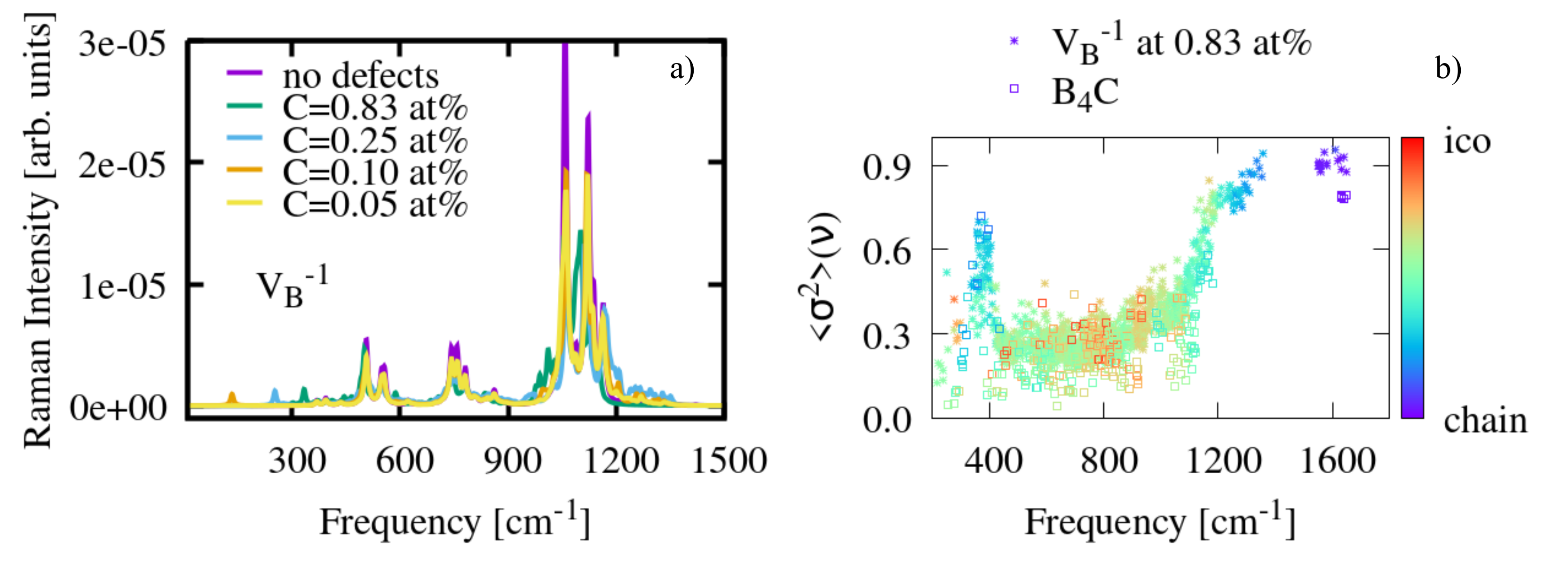}
  \caption{a) Theoretical Raman spectrum of the negatively-charged boron chain vacancy, at various defect concentrations. b) mode localisation compared to undefected bulk B$_4$C.}
  \label{FigVacancies}
\end{figure}

\begin{figure}
  \includegraphics[width=\columnwidth]{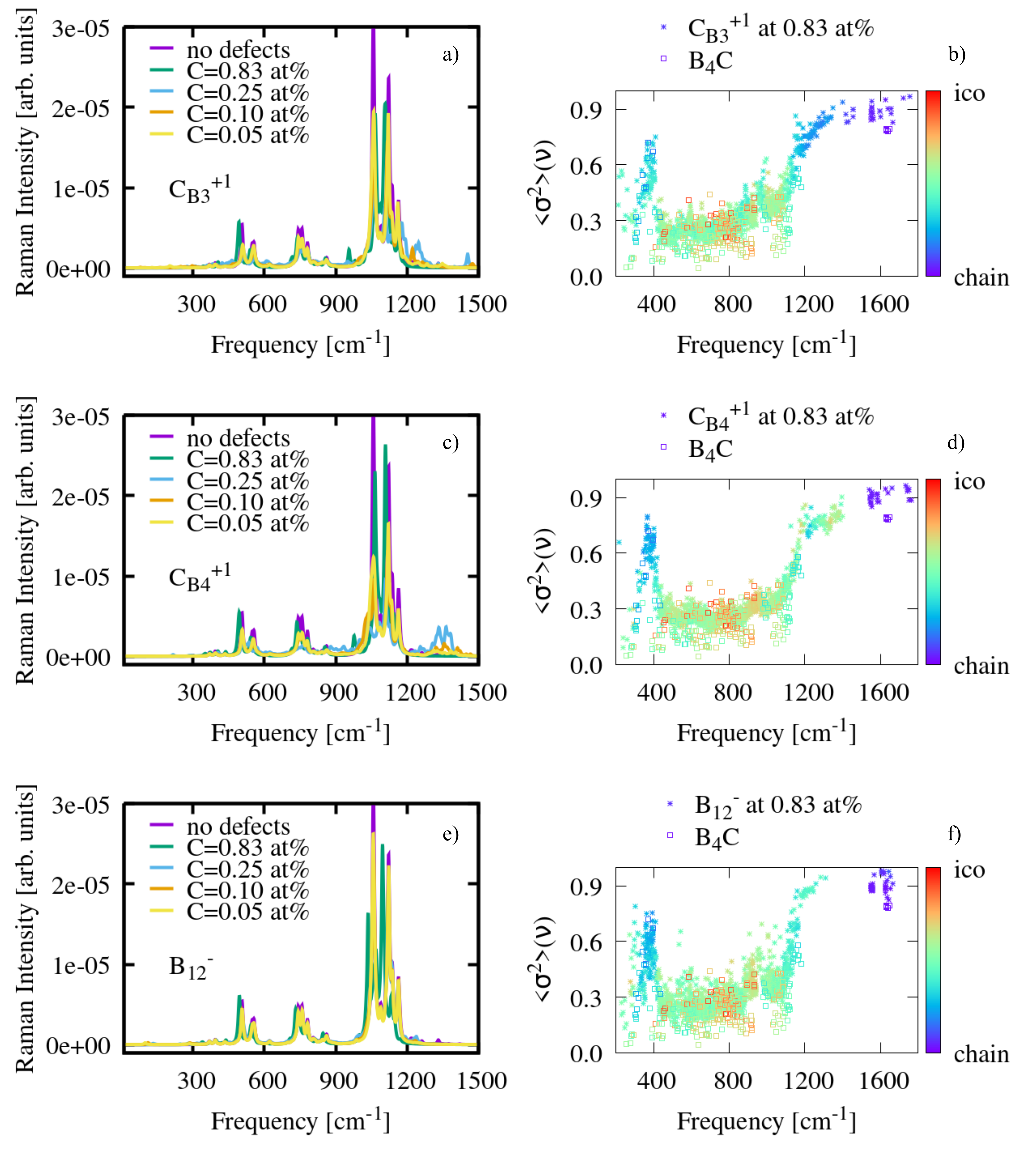}
  \caption{Theoretical Raman spectrum of two negatively-charged carbon antisites (panels a and c) and one positively-charged boron antisite (panel e), at various defect concentrations, and their mode localisation (panels b, d, f) compared to undefected bulk B$_4$C.}
  \label{FigAntisites}
\end{figure}

\begin{figure}
  \includegraphics[width=0.8\columnwidth]{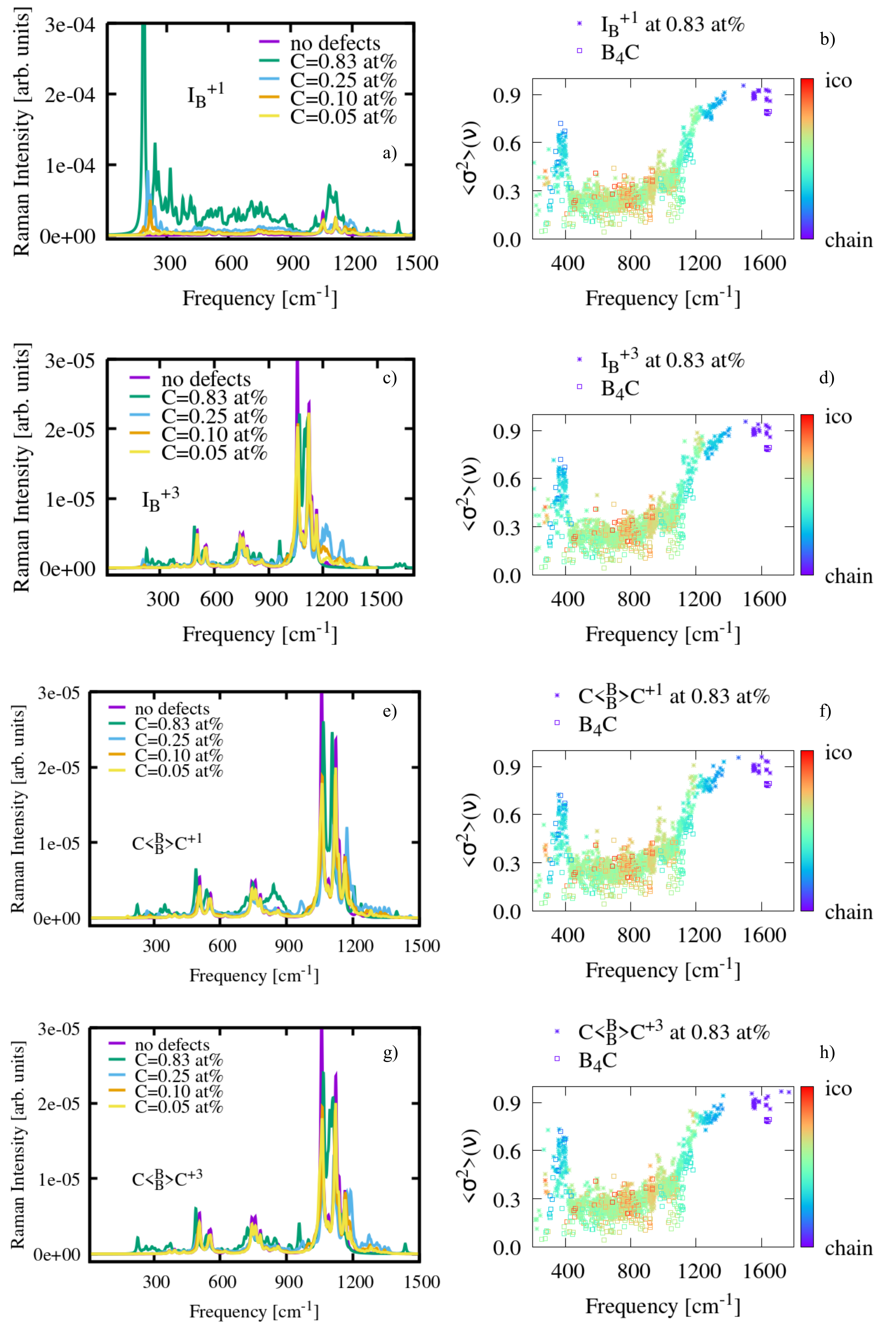}
  \caption{Theoretical Raman spectrum of two types of boron interstitials in charge states +1 and +3, at various defect concentrations (panels a,c,e,g), and mode localisation (panels b, d, f, h) compared to undefected bulk B$_4$C. The huge Raman activity of the boron interstitial in the Wyckoff position 2$c$ with charge +1 (panel a) is due to the small Fermi level range of stability of this defect, which questions the reliability of the underlying approximations of the present calculation (first order non-resonant Raman spectrum in the Placzek approximation).}
  \label{FigInterstitials}
\end{figure}

\begin{figure}
  \includegraphics[width=\columnwidth]{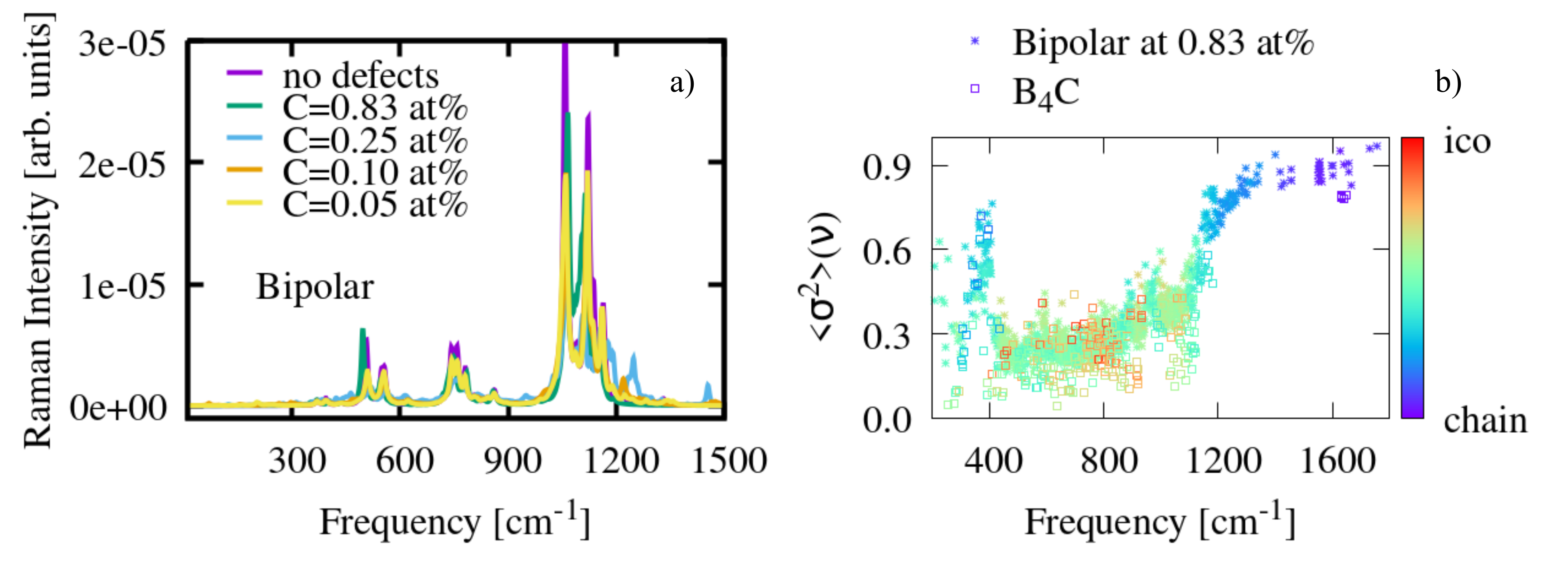}
  \caption{a) Theoretical Raman spectrum of the neutral bipolar complex, at various defect concentrations, and b) mode localisation of this defect compared to undefected bulk B$_4$C.}
  \label{FigBipolar}
\end{figure}

\begin{figure}
  \includegraphics[width=\columnwidth]{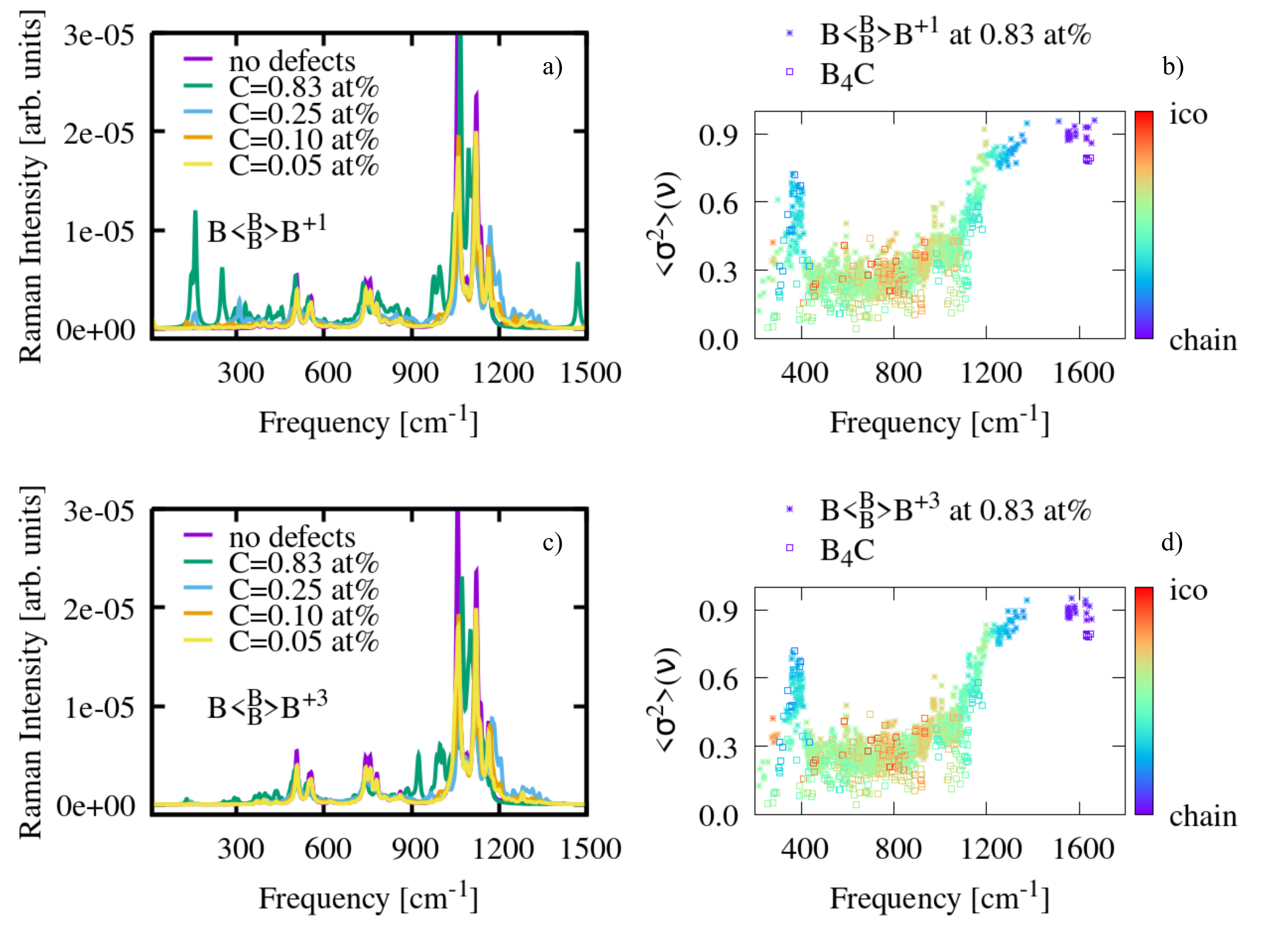}
  \caption{Theoretical Raman spectrum of the ~B$\langle^\text{B}_\text{B}\rangle$B blocks in charge states +1 (a) and +3 (c), at various defect concentrations, and mode localisation of this defect complex compared to undefected bulk B$_4$C (panels b and d). The charge transition level between these two charge states is at approximately 0.6~eV above the valence band top.}
  \label{FigBBBB}
\end{figure}

In figures \ref{FigVacancies}, \ref{FigAntisites}, \ref{FigInterstitials}, \ref{FigBipolar} we present the random Raman spectra obtained for supercells of B$_4$C containing various types of point defects: a singly negative boron chain vacancy  (Fig.\ref{FigVacancies}), singly positive boron antisites in positions B$_3$ and B$_4$ and singly negative icosahedral carbon antisite (Fig.\ref{FigAntisites}), +3 and +1 boron interstitials (Fig.~\ref{FigInterstitials}) and the bipolar defect (Fig.~\ref{FigBipolar}).
For all the mentioned defects we show the spectrum corresponding to various defect concentrations and a localisation chart. We performed a direct DFPT calculation for each of the defects, inserted in a 2$\times$2$\times$2 supercell (120 atoms) of the (B$_{11}$C)C-B-C 15-atom unit cell; this corresponds to a defect concentration of 0.83 atomic \%;  the spectra attributed to lower defect concentrations were obtained through an extrapolation procedure recently developed~\cite{roma_modeling_2019}. The localisation chart shows the localisation of vibrational modes (Raman active or not) for the supercell with the defect and for the perfect bulk B$_4$C.

When analysing Raman spectra of materials containing defects, in particular irradiated materials, one question frequently arises concerning the features of the Raman spectrum that appear only when defects are present: do they correspond to vibrational modes specifically associated to the defect, or are they vibrational modes already present in the density of states of the undefected crystal which became Raman active due to loss of symmetries? Saying that a mode is a ``defect mode'' implies that, first, this mode is not present in the undefected crystal and, second, its displacement pattern clearly involves almost exclusively the first shells of neighbours of the defect.

Let us consider first the region of the spectrum encompassing only a few dozens of cm$^{-1}$ around 1000 cm$^{-1}$, which is still not fully understood in boron carbide. Several defects introduce (or activate), at least at the highest investigated defect concentration of 0.83~at\%, some features in the Raman spectrum which are not present in the spectrum of the perfect (B$_{11}$C)C-B-C structure.
In the case of the boron chain vacancy (Fig.~\ref{FigVacancies}b), some of the modes in that region are slightly more localised than the bulk modes in the same region, which are however present, and the localisation of these supposedly defect modes, with moderate icosahedral caracter, are nevertheless less localised than most bulk chain modes.

Carbon icosahedral antisites C$_{B3}$ and C$_{B4}$ (Fig.~\ref{FigAntisites}) both present small peaks at 0.83~at\%, which disappear already at 0.25~at\%. For boron interstitials (Fig.~\ref{FigInterstitials}) the peak is slightly more pronounced, but in all cases the mode localisation in the same region is comparable to the one seen for the vacancy. The spectrum of the bipolar defect (Fig.~\ref{FigBipolar}a) does not present any remarkable feature in this region, although the mode localisation is slightly enhanced, as for the other defects.

A further defect complex was investigated, the ~B$\langle^\text{B}_\text{B}\rangle$B block, which can be interpreted as a combination of two chain boron antisites and one boron interstitial. Such a structure is an important component of the OPO$_1$ boron-rich phase. We calculated it in charge states +1 and +3, as their charge transition level is located only 0.6~eV above the top of the valence band. In both cases (see Fig.~\ref{FigBBBB} panels a and c) a clear hump in the Raman spectrum is present around 1000~cm$^{-1}$ at the highest concentration, but rapidly disappears as the defect concentration lowers.

We conclude then that probably several defects contribute to the experimentally observed shoulder around 980-1000 cm$^{-1}$, but no specific defect mode can be clearly identified as responsible for this feature.

We should spend a word concerning the spectrum of the singly positive boron interstitial I$_B^{+1}$ (Fig. \ref{FigInterstitials}a), whose intensity strongly evolves with its concentration: this should probably be interpreted as a breakdown of the Placzek approximation, which requires that the energy of vibrational modes are much lower than the band gap of the system. As the +1/+2 charge transition level is relatively close to the +1/-2 transition level (see Supplemental Material~\cite{roma_see_2021}), this condition is not satisfied any longer, especially in the smallest supercell (largest defect concentration) where some slight dispersion of electronic defect levels can be expected. The large peak below 300~cm$^{-1}$ reminds of peaks that arise in the experimental spectrum only in specific conditions, and do certainly deserve further investigation in the future.

Several defects present weakly Raman active modes between 1200~cm$^{-1}$ and 1400~cm$^{-1}$, where no bulk modes are present, but they virtually disappear as soon as the defect concentration lowers below 0.25~at\%.

At higher frequencies, some defects present one or more peaks slighly below 1500~cm$^{-1}$ and up to 1600 cm$^{-1}$. Such features, frequently attributed to free carbon~\cite{guo_pressure-induced_2010}, might in fact be caracteristics of boron carbide itself when Raman selection rules are lifted. This infrared mode is the most intense one, characteristic of the antisymmetric chain stretching, in the undefected (B$_{11}$C)C-B-C structure (see Fig.~\ref{FigBulkLoc}b). Indeed, it seems to be activated,
and sometimes shifted, due to the presence of defects.

The two main theoretical peaks close to 1070~cm$^{-1}$ and 1110~cm$^{-1}$ are not resolved experimentally, despite the fact that the resolution of today's Raman spectrometers should be sufficient to resolve them.
The reason of this can be related to the lifetime of one or both peaks (or of the weaker intermediate peak around 1085~cm$^{-1}$), or it may be associated to the presence of defects. The vacancy and the pantograph interstitial (~C$\langle^\text{B}_\text{B}\rangle$C$^{+3}$) look like possible candidates, if their concentration is large enough, but given their formation energy, their presence would imply high defect supersaturation.

\subsubsection{Raman spectra of ground-state and metastable complex phases}
\label{RamanComplex}

In this subsection we consider the ``3+1'' neutral complex
and the OPO$_1$ and OPO$_2$~\cite{rasim_local_2018,jay_theoretical_2019} structures.

\begin{figure}
  \includegraphics[width=\columnwidth]{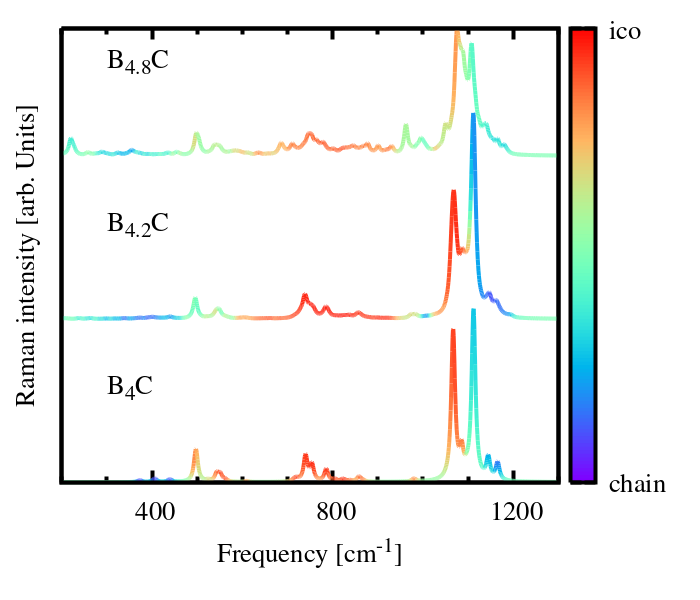}
  \caption{The Raman spectrum B$_4$C containing or not a ''3+1'' defect complex. The three spectra were obtained with a direct DFPT calculation. Those labelled B$_{4.8}$C and B$_{4.2}$C refer respectively to cells containing 406 and 121 atoms, built by inserting in B$_4$C a ''3+1'' defect complex. The former structure, with 19.2\% atomic carbon concentration, lies on the convex null in DFT-LDA and thus defines a stable phase at that stoichiometry. The corresponding spectrum is the one in the middle. The one at the top, corresponding to a higher concentration of the ''3+1'' defect complex, is shown for comparison. }
  \label{Fig3+1}
\end{figure}

Let us consider now the Raman spectrum of boron carbide containing ``3+1'' defects; we obtained it with a direct DFPT calculation both for the 2$\times$2$\times$2 (121 atoms) and 3$\times$3$\times$3 (406 atoms) supercells. The latter identifies the ``(3+1)@27'' ordered phase sitting on the convex hull, as discussed in section \ref{DefectComplexes}. The results are shown in Fig.~\ref{Fig3+1}. For the new phase ---built as an ordered arrangement of ``3+1'' defects, one every 27 (B$_{11}$C)C-B-C unit cells--- the spectrum is quite close to the one of perfect B$_4$C. The icosahedral character of the doublet around 500~cm$^{-1}$ seems to be lost, which is an indication that the eigenvectors of the concerned modes now have significant contribution from the boron interstitial displacement and/or nearby chains.

When the concentration of the ``3+1'' defect reaches 0.83 at.\% (one every 8 unit cells), the modifications are more pronounced, although the spectrum of B$_4$C is still clearly recognizable (top curve in Fig.~\ref{Fig3+1}). We note in particular a rise in the intensities in the region just below, and up to, 1000~cm$^{-1}$; The largest doublet around 1100~cm$^{-1}$ is also partly merged, two features that are observed in experiments. This fact is even more evident when taking into account geometrical effects, as shown in angular maps presented in the Supplemental Material~\cite{roma_see_2021}, Figure~S3.

What we simulate here is, of course, still a periodic arrangement of ``3+1'' defect complexes, due to periodic boundary conditions; as such it is not fully comparable to what would be an experimental situation where a given concentration of randomly distributed defects is present. Even from the point of view of the free energy, we remind that a random distribution of ``3+1'' defects would be stabilized by configurational entropy at finite temperature, at least, and to a less extent by the vibrational one\cite{jay_theoretical_2019}, which are both neglected here. The calculated spectra is nevertheless a hint that this defect complex, as other point defects mentioned before, might contribute to the shoulder at 1000~cm$^{-1}$ and to the merging of the two strongest peaks in the spectrum, especially considering its low formation energy in, even moderately, boron-rich conditions. This defect complex, both from the point of view of energetics at the DFT-LDA level and of its Raman spectrum, appears to be one of the missing concentrations, in the phase diagram between the carbon-rich (B$_{11}$C)C-B-C structure and the boron-rich OPOs phases.

\begin{figure}
  \includegraphics[width=0.6\columnwidth]{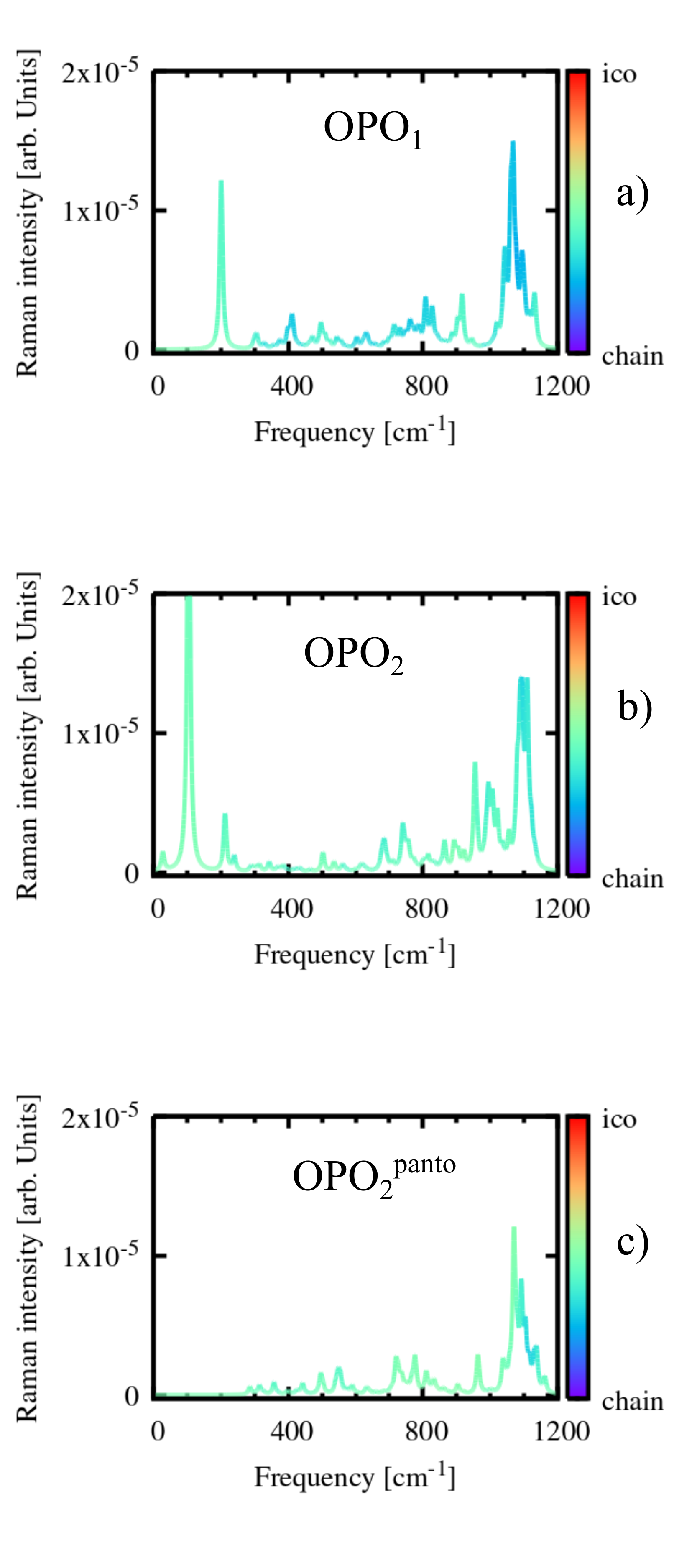}  
  \caption{Raman spectrum calculated for the two boron-rich phases called OPO$_1$ (a) and OPO$_2$ (b) and for the OPO$_2$ variant named OPO$_2^{panto}$ (c) featuring the ~C$\langle^\text{B}_\text{B}\rangle$C pantograph structure instead of $2c$ boron interstitials. The unit-cells of these phases all contain 414 atoms.}
  \label{FigOPOs}
\end{figure}

Finally, let us consider the spectrum calculated for the two OPO structures.
The calculated Raman spectrum for their 414 atoms unit cells is shown in Fig.~\ref{FigOPOs}. The color scale for the localisation of the modes is exactly the same as in the previous figure (Fig.~\ref{Fig3+1}) for the ``(3+1)@27'' phase. The different mode localisation is striking, because here the chain or icosahedral character of the modes is almost completely washed out. This is probably an effect of the large number of boron atoms that correspond to interstitials of the 15-atom unit cell. They are neither icosahedral nor chain atoms in the description 
and our analysis tool only involved icosahedral and chains atoms. For the OPOs structures it would certainly be useful to devise an extended classification of the atoms constituting the structure, in particular to include
the case of the~C-B-C$\cdots$B blocks of the OPO$_2$.

  The Raman active regions are much more extended than for bulk B$_4$C. Although some of the main features of the B$_4$C spectrum are still recognisable, the modifications are much deeper than in the spectra containing one defect at relatively low concentration. While the OPO$_2$ spectrum clearly shows fairly intense features in the 900-1000~cm$^{-1}$ region, the OPO$_1$ lacks such contribution. One might deduce that the experimental shoulder in that region is more likely due to boron interstitials in the Wyckoff position 2$c$, or maybe ~C$\langle^\text{B}_\text{B}\rangle$C pantograph blocks, than to~B$\langle^\text{B}_\text{B}\rangle$B structures present in the OPO$_1$ phase; however, as we have shown in Fig.~\ref{FigBBBB}a,c, the ~B$\langle^\text{B}_\text{B}\rangle$B defect complex, at a  concentration close to 1~at\% (corresponding to 19.2\% carbon), clearly shows such a contribution; the fact that this region is not Raman active in the calculated spectrum of the OPO$_1$ must then have a more complex explanation.
 
To conclude we recall the already mentioned low frequency broad doublet occurring around 300~cm$^{-1}$. Although some of the point defects studied, and some of the complex phases, show peaks below 400~cm$^{-1}$, in particular when boron interstitials are present (see Figs.~\ref{FigInterstitials}a,c,e,g, \ref{FigBBBB}c, \ref{Fig3+1}, and \ref{FigOPOs}a,b) we cannot propose any solid identification on the basis of our calculations. Given the sensitivity of such features to the excitation frequency and power one could imagine that resonant effects should be taken into account~\cite{Yan_B4Cdeamorph_PRL2009,guo_pressure-induced_2010}, which are not included in the present approach. Possibly, a combination of defects and resonant effects involving their electronic energy levels located in the band gap should be investigated in more detail.
  


\section{Summary and conclusions}
\label{Conclu}
In this paper we addressed several aspects of the first order Raman spectrum of boron carbide in particular in connection with the influence of point defects on the Raman spectrum, mainly, but not exclusively, from a theoretical perspective.
First, the analysis of the Raman spectrum of the well known (B$_{11}$C)C-B-C structure, accepted as the most stable structure for the stoichiometric ratio B$_4$C, shows that the intensity of some peaks is strongly dependent on the experimental geometry. A comparison of an experimental spectrum collected along a line connecting two grains of different orientation, confirms this fact.

Second, we propose to combine a consistent definition of the localisation of vibrational modes and a labelling of atoms (chain, icosahedral or neither of them) to deepen the analysis of the active Raman peaks.
Our results confirm some previous labelling of modes and provide further insight into the localisation of bulk and defect modes.

Third, we study the most probable point defects of carbon-rich boron carbide and their influence on the Raman spectrum. After summarizing the results of a systematic study of defect energetics, we analyse the spectra of defect containing supercells corresponding to relatively low concentrations and we extrapolate to even lower concentrations thanks to our embedding tool. Several defects show a contribution in the region of the up to now unexplained shoulder near 1000~cm$^{-1}$, confirming previous speculations that this feature could be induced by defects. However, our results show that it is not a single defect that can be considered responsible for such a feature, but rather a variety of point defects. Moreover, the corresponding activated peaks do not seem to be specifically defect modes, because they seem to involve a significant displacement of several atoms, i.e., they show a fairly collective character.

The fourth and last contribution of this work, is to analyse the Raman spectra of ground state phases,  and of the metastable ones, and notably of the Ordered Partial Occupations (OPO) structures we have recently proposed as ground state structures in the boron-rich domain. Furthermore, we propose that the ground state at 19.2~\% atomic carbon concentration is given, in DFT-LDA, by the phase with 406 atoms unit-cell containing the defect complex that we dub ``3+1''. The latter, formed by a boron interstitial atom and three (B$_{12}$) icosahedra, reveals a stabilization by the Coulomb interaction between the constituent defects. The insertion of this structure in an otherwise bulk environment leads to a very stable atomic configuration, reasonably representing boron carbide in the large region of stoichiometry between the OPO$_2$ structure (boron-rich) and the B$_4$C stoichiometry (carbon-rich). Convincingly, the analysis of the Raman spectrum of the ``3+1'' defect shows a contribution to the 1000~cm$^{-1}$ shoulder and to the merging of the two main peaks at higher frequency, another aspect of experimental spectra.





\section{Acknowledgements}
We acknowledge fruitful discussions with Olivier Hardouin Duparc, Jelena Sjakste, Romuald B\'ejaud, Dominique Gosset and Aur\'elien Jankowiak and we thank the last two for preparing the samples.
Financial support from the Programme NEEDS-MAT{\'E}RIAUX, which initiated this work, as well by the French DGA and the DIM SIRTEQ (région Île de France) as is gratefully aknowledged.
This work was granted access to HPC resources by the Partnership for Advanced Computing in Europe (PRACE Project No. 2019204962), by the French HPC centers GENCI-CINES and GENCI-TGCC (Projects 2020A0010906018s, 2210, and allocation by CEA-DEN) and by \'Ecole Polytechnique through the LLR-LSI project. 

\end{document}